\documentclass[aps,prb,twocolumn,showpacs,preprintnumbers,amsmath,amssymb,superscriptaddress]{revtex4}

\usepackage{graphicx}% Include figure files
\usepackage{dcolumn}% Align table columns on decimal point
\usepackage{bm}% bold math

\begin{document}

\title{Large enhancement of thermoelectric effects in a double quantum dot system due
to interference and Coulomb correlation phenomena}

\author{Piotr Trocha}
\email{ptrocha@amu.edu.pl}\affiliation{Faculty of Physics, Adam
Mickiewicz University, 61-614 Pozna\'n, Poland}

\author{J\'ozef Barna\'s}
\email{barnas@amu.edu.pl} \affiliation{Faculty of Physics, Adam
Mickiewicz University, 61-614 Pozna\'n, Poland}
\affiliation{Institute of Molecular Physics, Polish Academy of
Sciences, 60-179 Pozna\'n, Poland}

\date{\today}

\begin{abstract}

Thermoelectric effects in a double quantum dot system coupled to
external magnetic/nonmagnetic leads are investigated
theoretically. The basic thermoelectric transport characteristics,
like thermopower, electronic contribution to heat conductance, and
the corresponding figure of merit, have been calculated in terms
of the linear response theory and Green function formalism in the
Hartree-Fock approximation for Coulomb interactions. An
enhancement of the thermal efficiency (figure of merit $ZT$) due
to Coulomb blockade has been found. The magnitude of ZT is further
considerably enhanced by quantum interference effects. Both the
Coulomb correlations and interference effects lead to strong
violation of the Wiedemann-Franz law. The influence of
spin-dependent transport and spin bias on the thermoelectric
effects (especially on Seebeck and spin Seebeck effects) is also
analyzed.


\end{abstract}

\maketitle

\section{Introduction}\label{Sec:1}

Energy conversion based on thermoelectric properties of
solid-state  materials has been attracting recently much
attention, especially in the case of nanoscale systems. Although
thermoelectric phenomena have been known for long time, their
efficiency in bulk materials was relatively low, and therefore
systems of high thermoelectric efficiency are greatly desired. The
efficiency is usually measured by the dimensionless thermoelectric
figure of merit $ZT$, $ZT=\sigma S^2T/\kappa$. Here, $T$ stands
for the temperature, $\sigma$ is the charge conductivity, $S$ is
the Seebeck coefficient (thermopower), and $\kappa$ is the thermal
conductivity which generally includes the phonon and electronic
contributions. From this formula it is clear that to enhance the
efficiency of a thermoelectric device one has to somehow enlarge
the thermopower $S$ and electric conductivity $\sigma$, and reduce
the thermal conductivity $\kappa$. However, this task is difficult
to be realized in conventional materials which obey the
Wiedemann-Franz law.~\cite{wied-franz} Moreover, the Seebeck
coefficient becomes decreased as the charge conductance increases
due to the Mott relation.~\cite{mott} Typically observed values of
$ZT$ are smaller than  one, $ZT\lesssim 1$,  and this is why
thermoelectric materials have not been widely used in commercial
applications.

It has turned out recently, that this drawback of  bulk materials
can be overcome in nanoscale systems. Hick \emph{et al.}~\cite{hick}
have predicted an increase of the figure of merit as the dimension
of the system is reduced. Now, it is well known that the
thermoelectric properties of nanoscale systems are strongly
affected by quantum confinement (level quantization) and Coulomb
blockade
effects.~\cite{beenak,blanter,turek,koch,kubala1,zianni,zhangxm}
These effects can  lead to violation of the Wiedemann-Franz law
and failure of the Mott relation
.~\cite{kubala2,murphy} Moreover, the thermal conductance of low
dimensional systems is rather small, which allows to reach rather
high values of $ZT$.~\cite{chen} Owing to these properties of
nanoscale structures, which are promising from the application
point of view, interest in thermoelectric phenomena revived in
recent years. Accordingly, thermoelectric properties of quantum
dots (QDs), quantum wires, molecules and silicon nanojunctions
have been investigated both
experimentally~\cite{reddy,hochbaum,baheti,boukai,schwab} and
theoretically.~\cite{rego,dubi,swirkowicz,markussen,galperin,lunde,segal,pauly,LiuNano09,bergfield}
Oscillations of the thermopower\cite{beenak,blanter} and
  thermal conductance~\cite{zianni,kubala2} with
external gate voltage have been reported in thermoelectric
transport through Coulomb islands. Moreover, it has been shown
that spin correlations can strongly influence the thermoelectric
effects in the Kondo
regime.~\cite{boese,dong,krawiec,sakano,kim,scheibner,franco,yoshida,costi}
Recently, an increase of the thermal efficiency has been reported
in a multilevel QD for temperatures much smaller than the intradot
charging energy.~\cite{swirkowiczJP,liu,kuo}
Furthermore, the influence of ferromagnetic leads (spin polarized
current) on thermoelectric transport properties in a single-level
QD has been also investigated in different coupling
regimes.~\cite{dubi,swirkowicz} In turn, Hatami \emph{et al.} have
analyzed the Peltier and Seebeck phenomena in magnetic
multilayered structures and shown that the thermoelectric effects
significantly depend on the relative alignment of magnetizations
in neighboring magnetic layers.~\cite{hatami} For a more
comprehensive description of various thermoelectric transport
properties in nanostructures we refer to recent review
articles.~\cite{Snyder,wangRev,dubiRev}

However, the thermoelectric phenomena in multiple quantum dot
systems are rather unexplored.~\cite{zhangdqd,chiPLA11,swirkowiczPRB11,orellanaPre11} Especially, double
quantum dot systems may exhibit some new thermoelectric phenomena
as they reveal a variety of different interference effects,
including Fano and Dicke resonances~\cite{trocha07,trochaJNN} or
Aharonov-Bohm oscillations.~\cite{loss} The Fano effect appears due
to quantum interference of waves resonantly transmitted through a
discrete level and those transmitted nonresonantly through
continuum of states. The  effect was first observed as an
asymmetric line shape of atomic emission spectra, but it also
appears in electronic transport through QD systems. In the case of
double QDs, the Fano effect originates from the quantum
interference of electron waves transmitted coherently through the
weakly coupled state to the leads and those transmitted through
the state which is strongly coupled to the external
electrodes.~\cite{guevara,trocha07}
Very recently Liu \emph{et al.} have investigated thermoelectric
effects in parallel double QDs attached to two metallic leads, and
with a magnetic flux threading the quantum dot
device.~\cite{liuJAP} They arrived to the conclusion that the
figure of merit $ZT$ can be enhanced in the vicinity of the Fano
resonance. Similar conclusion also follows from a recent
paper~\cite{karlstrom}, where the influence of electron
interference in a two-level system on the maximum thermoelectric
power is analyzed.

Recently, the spin voltage generated by a temperature gradient has
been observed experimentally in a metallic magnet.~\cite{uchida}
This novel phenomenon, called spin Seebeck effect, offers a new
way for generation of a pure spin current without accompanying
charge current. The discovery of spin Seebeck effect has
stimulated both experimental and theoretical interest in the
so-called spin caloritronics. Altough the spin caloritronics is as
old as spin electronics~\cite{johnsonPRB87}, it has been poorly
investigated and has remained dormant for many years -- except of
a few experimental works on the thermoelectric properties of
magnetic multilayers in the current-in-plane geometry.~\cite{shi}
More recently spin-dependent heat and charge transport in magnetic
multilayers was studied experimentally in the
perpendicular-to-plane geometry.~\cite{gravier,gravier2}
As concerns spin thermoelectric effects in quantum dot systems, these are rather poorly explored although interest in them has been growing recently.~\cite{swirkowicz,dubiPRB09,LiuJAP11,rejecPre11}

In this paper we consider thermoelectric effects in the system
consisting of parallel coupled quantum dots embedded between
metallic (normal or ferromagnetic) leads. In contrast to
Ref.[\onlinecite{liuJAP}], we include the Coulomb correlations.
The latter are long range interactions and therefore can not be
excluded in realistic systems. Moreover, one can expect a
significant enhancement of the thermal efficiency in the Coulomb
blockade regime. The basic thermoelectric characteristics are
derived using nonequilibrium Green function method. The Coulomb
correlations are taken in the Hartree-Fock approximation and thus
the formalism is relevant for temperatures higher than the
specific temperature associated with the Kondo effect (Kondo
temperature). Additionally, we investigate the spin Seebeck
effect, too.  In Section~\ref{Sec:2} we present the model of a
double quantum dot system  attached to ferromagnetic/nonmagnetic
leads and describe basic thermoelectric phenomena to be analyzed.
In Sec.~\ref{Sec:3} we present the corresponding numerical
results. Finally, Sec.~\ref{Sec:4} includes summary and general
conclusions.

\section{Theoretical description}\label{Sec:2}

\subsection{Model}

We consider two single-level quantum dots which are attached to
normal/magnetic metallic leads. Since the inter-dot Coulomb
interaction is small in comparison to the intra-dot one, the
former will be omitted in the following. The system is then
described by Hamiltonian of the form
\begin{align}\label{Eq:1}
\hat{H}&=\sum_{\mathbf{k}\beta\sigma}\limits
\varepsilon_{\mathbf{k}\beta \sigma}c^{\dagger}_{\mathbf{k}\beta
\sigma} c_{\mathbf{k}\beta \sigma}+\sum_{\mathbf{k}\beta}\limits\sum_{i\sigma}
   \limits (V_{i\mathbf{k}\sigma}^\beta c^\dag_{\mathbf{k}\beta\sigma}d_{i\sigma}+\rm
   h.c.)
\notag \\
&+\sum_{i\sigma}\limits\varepsilon_{i\sigma}d^\dag_{i\sigma}d_{i\sigma}-
   t\sum_\sigma\limits(d^\dag_{1\sigma}d_{2\sigma}+h.c.)
    + \sum_i\limits
   U_in_{i\sigma}n_{i\bar{\sigma}}
\end{align}
where the first term describes the left ($\beta= L$) and right
($\beta= R$) leads in the non-interacting quasi-particle
approximation. Here, $c^{\dagger}_{\mathbf{k}\beta\sigma}$
($c_{\mathbf{k}\beta\sigma}$) is the creation (annihilation)
operator of an electron with the wave vector $\mathbf{k}$ and spin
$\sigma$ in the electrode $\beta$, whereas
$\varepsilon_{\mathbf{k}\beta \sigma}$ denotes the corresponding
single-particle energy. The second term of Hamiltonian
(\ref{Eq:1}) describes spin conserving electron tunneling between
the leads and $i$th dot ($i=1,2$), with
$V_{i\mathbf{k}\sigma}^\beta$ being the relevant matrix elements.
The last three terms describe two quantum dots, where
$\varepsilon_{i\sigma}$ denotes the discrete energy level of the
$i$-th dot, $n_{i\sigma}=d^\dag_{i\sigma}d_{i\sigma}$ is the
corresponding  particle number operator, $t$ is the inter-dot
hopping parameter (assumed real and independent of the electron
spin orientation), whereas $U_i$ is the Coulomb energy
corresponding to double occupation of the $i$-th dot ($i=1,2$).

The dot-leads coupling is described usually by the width functions
$\Gamma_{ij\sigma}^\beta = 2\pi \rho_\beta^\sigma V_{i}^{\beta}
V_{j}^{\beta\ast}$, where $\rho_{\beta}^\sigma$ is the density of
states in the lead $\beta$ for spin $\sigma$.
$\Gamma_{ij\sigma}^{\beta }$ describes the spin-dependent
hybridization of the dots' levels ($i,j=1,2$) and states of the
$\beta$th lead. For the sake of simplicity we assume that the
couplings are constant within the electron band;
$\Gamma^\beta_{ij\sigma}(\varepsilon)=\Gamma^\beta_{ij\sigma}={\rm
const}$ for $\varepsilon\in\langle-D,D\rangle$, and
$\Gamma^\beta_{ij\sigma}(\varepsilon)=0$ otherwise, where $2D$
denotes the electron band width.

The coupling parameters can be written in a matrix form as
\begin{equation}\label{Eq:2}
  \mathbf{\Gamma}^\beta_\sigma = \left(
\begin{array}{cc}
  \Gamma^\beta_{11\sigma} & q_\beta\sqrt{\Gamma^\beta_{11\sigma}\Gamma^\beta_{22\sigma}} \\
 q_\beta\sqrt{\Gamma^\beta_{11\sigma}\Gamma^\beta_{22\sigma}} & \Gamma^\beta_{22\sigma}
\end{array}
\right),
\end{equation}
for $\beta =L,R$. To take into account various interference
effects leading to suppression of the nondiagonal terms, we have
introduced the parameters $q_{\rm L}$ and $q_{\rm R}$. These
parameters are generally complex, and $ 0\le \vert q_\beta \vert
\le 1$.
The spin-dependent coupling strengths can be then written as;
$\Gamma^L_{11\sigma}=\Gamma_L(1\pm p_L)$,
$\Gamma^L_{12\sigma}=\Gamma^L_{21\sigma}=q_L\Gamma_L\sqrt{\alpha}(1\pm
p_L)$, $\Gamma^L_{22\sigma}=\alpha\Gamma_L(1\pm p_L)$,
$\Gamma^R_{11\sigma}=\alpha\Gamma_R(1\pm \delta p_R)$,
$\Gamma^R_{12\sigma}=\Gamma^R_{21\sigma}=q_R\Gamma_R\sqrt{\alpha}(1\pm
\delta p_R)$, and $\Gamma^R_{22\sigma}=\Gamma_R(1\pm \delta p_R)$
with $\delta=1$ for parallel magnetic configuration and
$\delta=-1$  for the antiparallel one. The upper sign refers here
to $\sigma=\uparrow$, while the lower sign refers to
$\sigma=\downarrow$. Apart from this, $p_\beta$ ($\beta=L,R$) is
the polarization factor of the $\beta$-th lead, $\Gamma_L$ and
$\Gamma_R$ are the coupling constants, and $\alpha$ takes into
account difference in the coupling of a given electrode to the two
dots. For $\alpha=0$ the system becomes reduced to two QDs
connected in series.

\subsection{Thermoelectric phenomena}

Consider first the situation when chemical potentials of the leads
are independent of spin orientation (no spin accumulation). The
charge current $J$, spin current $J^s$,  and heat current of
electronic origin $J^Q$, flowing from the left lead to the right
one, can be determined from the following formula:
\begin{equation}\label{Eq:3}
\left(
  \begin{array}{c}
    J \\
    J^s \\
    J^Q \\
  \end{array}
\right)
=
\frac{1}{h}\sum_{\sigma}\limits\int d\varepsilon\left(
  \begin{array}{c}
     e \\
      \hat{\sigma} \hbar/2  \\
     \varepsilon-\mu_L\\
  \end{array}
\right)[f_L(\varepsilon)-f_R(\varepsilon)]T_{\sigma}(\varepsilon),
\end{equation}
where
$f_\beta(\varepsilon)=\{\exp[(\varepsilon-\mu_\beta)/k_BT_\beta]+1\}^{-1}$
is the Fermi-Dirac distribution function for the lead $\beta$ with
$\mu_\beta$ and $T_\beta$ denoting the corresponding chemical
potential and temperature, and $k_B$ standing for the Boltzmann
constant. Furthermore, $T_{\sigma}(\varepsilon)$ is the sum of
transmission coefficients through the two conducting channels
associated with two dots for carriers with spin $\sigma$ ($\sigma
=\uparrow, \downarrow$), and $\hat\sigma =1$ for $\sigma
=\uparrow$ and $\hat\sigma =-1$ for $\sigma =\downarrow$.
$T_{\sigma}(\varepsilon)$ can be expressed by the Fourier
transforms of retarded ($\mathbf{G}^r_\sigma$) and advanced
($\mathbf{G}^a_\sigma$) Green functions of the dots and by the
coupling matrices  $\mathbf{\Gamma}^{\beta}_\sigma$ ($\beta=L,R$);
$T_{\sigma}(\varepsilon)=Tr[\mathbf{G}^a_\sigma\mathbf{\Gamma}^R_\sigma\mathbf{G}^r_\sigma\mathbf{\Gamma}^L_\sigma]$.
The Green functions have been  calculated by the equation of
motion technique in the  Hartree-Fock approximation for the
Coulomb term.~\cite{trocha07}

In the linear response regime, Eq.(\ref{Eq:3}) can be transformed
to the following formulas for charge, spin, and heat currents:
\begin{equation}\label{Eq:4}
J=\sum_\sigma J_\sigma \equiv e \sum_\sigma L_{0\sigma}
 \delta \mu+\frac{e}{T}\sum_\sigma L_{1\sigma}\delta  T,
\end{equation}
\begin{equation}\label{Eq:5}
J^s=\sum_\sigma J^s_\sigma =\frac{\hbar}{2}\sum_\sigma
\hat{\sigma} L_{0\sigma}\delta \mu+\frac{\hbar}{2T}\sum_\sigma
\hat{\sigma} L_{1\sigma}\delta  T,
\end{equation}
\begin{equation}\label{Eq:6}
J^Q=\sum_\sigma J^Q_\sigma \equiv \sum_\sigma L_{1\sigma}\delta
\mu +\frac{1}{T} \sum_\sigma L_{2\sigma}\delta  T,
\end{equation}
where $\delta  T$ is the difference in temperatures of the leads,
and $\delta \mu=e\delta  V$ with $\delta  V$ being the voltage
drop between the two electrodes. We remind, that according to our
assumption, the chemical potential and temperature of the left
electrode are $\mu +\delta  \mu$ and $T +\delta  T$, respectively,
whereas of the right electrode are $\mu$ and $T$. Note, that in
the linear response regime both $\delta \mu$ and $\delta T$ tend
to zero. In Eqs.~(\ref{Eq:4}) to (\ref{Eq:6}) $L_{n\sigma}$
($n=0,1,2;\sigma =\uparrow, \downarrow$) are the integrals of the
form
\begin{equation}\label{Eq:7}
L_{n\sigma}=-\frac{1}{h}\int  d\varepsilon
(\varepsilon-\mu)^n\frac{\partial
f}{\partial\varepsilon}T_{\sigma}(\varepsilon).
\end{equation}

The thermopower $S$ is defined as the ratio of the voltage
drop $\delta  V$ generated by the temperature difference $\delta
T$, $S=\delta  V/\delta  T$, taken in the absence of charge
current, $J=0$. Thus, taking into account  Eq.(\ref{Eq:4}) and
Eq.(\ref{Eq:6}), one obtains the following formula for the Seebeck
coefficient:
\begin{equation}\label{Eq:8}
S\equiv\left[\frac{\delta  V}{\delta
T}\right]_{J=0}=-\frac{1}{eT}\frac{\sum_\sigma
L_{1\sigma}}{\sum_\sigma L_{0\sigma}}.
\end{equation}
Similarly, the charge  and spin  conductances, $G$ and $G^s$, can
be expressed in terms of the integrals (\ref{Eq:7}) as~\cite{ziman}
\begin{equation}\label{Eq:9}
G=e^2\sum_\sigma L_{0\sigma},
\end{equation}
\begin{equation}\label{Eq:10}
G^s=\frac{e\hbar}{2}\sum_\sigma \hat{\sigma}L_{0\sigma},
\end{equation}
while the thermal conductance can be written as
\begin{equation}\label{Eq:11}
\kappa=\frac{1}{T}\left(\sum_\sigma L_{2\sigma}-\frac{[\sum_\sigma
L_{1\sigma}]^2} {\sum_\sigma L_{0\sigma}}\right).
\end{equation}
Note that the thermal conductance is determined on the condition
of vanishing charge current. Thus, to determine the basic
thermoelectric parameters we need to find all the relevant
integrals, $L_{0\sigma}$, $L_{1\sigma}$, and $L_{2\sigma}$.

Consider now a more general situation, when spin accumulation in
the external leads becomes relevant, eg. due to long spin
relaxation time or when an external spin dependent bias is applied
to the system. In such a case we have to take into account spin
splitting of the chemical potential in the leads. In a general
case, temperature may also be different in different spin
channels. However, we neglect this assuming $T$ independent of
$\sigma$. This may be justified as the energy relaxation time is
much shorter than the spin relaxation one. Equations~(\ref{Eq:3})
take now the form
\begin{equation}\label{Eq:12}
\left(
  \begin{array}{c}
    J \\
    J^s \\
    J^Q \\
  \end{array}
\right) = \frac{1}{h}\sum_{\sigma}\limits\int d\varepsilon\left(
  \begin{array}{c}
     e \\
     \hat{\sigma} \hbar/2  \\
     \varepsilon-\mu_L\\
  \end{array}
\right)[f_{L\sigma}(\varepsilon)-f_{R\sigma}(\varepsilon)]T_{\sigma}(\varepsilon),
\end{equation}
where the spin dependence of the Fermi distribution function is
now indicated explicitly. Thus, charge, spin and heat currents can
still be expressed in the form of Eqs (\ref{Eq:4}) to
(\ref{Eq:6}), but with $\delta\mu$ being explicitly dependent on
$\sigma$, $\delta\mu\to \delta\mu_\sigma$, and $L_{n\sigma}$
including now derivative of spin dependent Fermi distribution
function.

Since the bias is now spin dependent, the difference in chemical
potentials, $\delta\mu_\sigma$, in the spin channel $\sigma$ can
be written as
\begin{equation}\label{Eq:13}
\delta\mu_\sigma = e\delta V_\sigma =e(\delta V+\hat{\sigma}
\delta V^s),
\end{equation}
where $\delta V$ is the voltage bias and $\delta V^s$ is the spin
bias~\cite{swirkowicz}. Of curse, $\delta V^s=0$ in the absence of
spin accumulation.

Charge and spin currents can be written as $J=G\delta V +(2e/\hbar
)G^s\delta V^s$, where $G$ and $G^s$ are the linear charge and
spin conductances defined above (see Eqs (9) and (10)). In turn,
the spin current can be written as $J^s=G^s\delta
V+(\hbar/2e)G\delta V^s$.

The thermopower can be now calculated on the condition of
vanishing simultaneously both spin current and charge current, or
equivalently on the condition of vanishing charge current in each
spin channel. As a result, one can define spin dependent
thermopower as
\begin{equation}\label{Eq:16}
S_\sigma =\frac{\delta V_\sigma}{\delta T}= -\frac{L_{1\sigma}}{
eTL_{0\sigma}}.
\end{equation}
Equivalently,  one can define spin thermopower $S^s$
\begin{equation}\label{Eq:17}
S^s =\frac{\delta V^s}{\delta T}= \frac{1}{2}(S_\uparrow
-S_\downarrow)=-\frac{1}{2eT}\left(\frac{L_{1\uparrow}}{
L_{0\uparrow}}-\frac{L_{1\downarrow}}{ L_{0\downarrow}}\right)
\end{equation}
in addition to the usual thermopower
\begin{equation}\label{Eq:18}
S =\frac{\delta V}{\delta T}= \frac{1}{2}(S_\uparrow
+S_\downarrow)=-\frac{1}{2eT}\left(\frac{L_{1\uparrow}}{
L_{0\uparrow}}+\frac{L_{1\downarrow}}{ L_{0\downarrow}}\right).
\end{equation}
In turn, the heat conductance is then given by
\begin{equation}\label{Eq:19}
\kappa=\sum_\sigma \kappa_\sigma \equiv
\frac{1}{T}\sum_\sigma\left(L_{2\sigma}-\frac{L_{1\sigma}^2}{L_{0\sigma}}\right).
\end{equation}
In the following we use the above formulas to calculate numerically
the relevant thermoelectric coefficients.

\section{Numerical results}\label{Sec:3}

For numerical calculations we assume spin degenerate and equal dot
levels, $\varepsilon_{i\sigma}=\varepsilon_0$ (for $i =1,2$ and $\sigma
=\uparrow ,\downarrow$). We also assume similar magnetic
electrodes, $p_L=p_R\equiv p$, and symmetrical couplings,
$\Gamma_{\rm L}=\Gamma_{\rm R}\equiv\Gamma$. Typical experimental
values of the dot-lead coupling may vary from a few
microelectronovolts to a few milielectronovolts.~\cite{wiel} In the
following it is convenient to relate energy quantities to the
dot-lead coupling. To have unique energy unites, we write the
parameter $\Gamma$ as $\Gamma = \gamma\Gamma_0$, where $\Gamma_0$
is constant and will be treated as the energy unit (the energy
quantities will be related to $\Gamma_0$). In turn, $\gamma$ is a
dimensionless parameter that describes strength of the dot-lead
coupling in terms of $\Gamma_0$.

The case of nonmagnetic leads is treated as a special case
corresponding to $p=0$. The parameters $q_L$ and $q_R$ are assumed
to be real and equal, $q_L=q_R=q$. In turn, for the parameter
$\alpha$ we assume $\alpha=0.15$, which indicates that there is a
relatively large difference in the coupling of a given electrode
to the two dots. This parameter, however, can be tuned by external gate voltages.
For the sake of simplicity we also assume the
same Coulomb parameters for the two dots, $U_1=U_2=U$.

In a general case, the electrochemical potentials of the leads may
be spin dependent. In other words, in addition to the usual
voltage bias one may also consider spin bias. The latter may
appear when spin relaxation in the leads is slow so a spin
accumulation may appear due to the spin bottle-neck effect (or
simply when a spin bias is applied intentionally). This leads to a
number of spin thermoelectric effects. We start from the simpler
case when the spin relaxation in the leads is sufficiently fast to
neglect spin accumulation, so the leads' electrochemical
potentials are independent of the electron spin orientation.

\subsection{Absence of spin bias}\label{Sec:3a}

Since the thermoelectric effects considered in the manuscript
depend on spin polarization of the leads, we consider first the
situation with nonmagnetic electrodes, and then proceed to
magnetic leads.

\subsubsection{Nonmagnetic leads: $p=0$}\label{Sec:3aa}

The effects considered in this paper depend significantly on the
ratio of thermal energy and coupling strength of the dots to
electrodes. Let us consider first the low temperature regime,
$k_BT\ll\Gamma$, or in our notation, $k_BT/\Gamma_0\ll\gamma$.  We
have calculated the basic thermoelectric characteristics, like
thermopower $S$, heat conductance $\kappa$, charge conductance
$G$, and figure of merit $ZT$. Although we consider the  regime
$k_BT\ll\Gamma$, the temperature assumed here is higher than the
corresponding Kondo temperature, so we do not take into account
the Kondo correlations. In this transport regime one finds
generally  $ZT\ll 1$.~\cite{murphy} However, this can be changed by
quantum interference effects.

\begin{figure}[t]
\includegraphics[width=0.46\textwidth,angle=0]{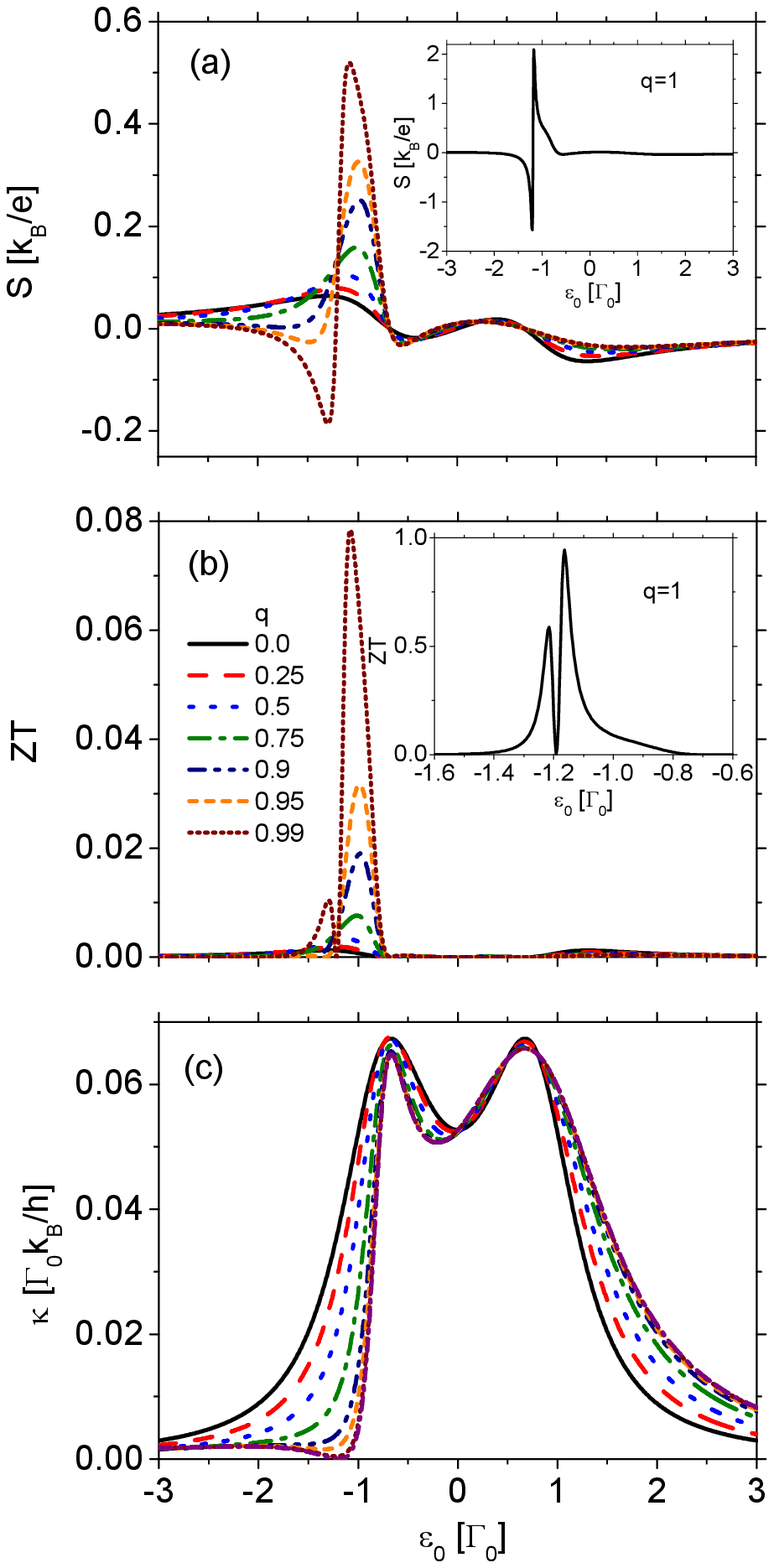}
\caption{\label{Fig:1} (color online) Thermoelectric coefficients:
(a) thermopower, (b) the figure of merit, (c) thermal conductance,
calculated as a function of the dots' levels energy for indicated
values of the parameter $q$. The other parameters are $k_BT/\Gamma_0
=0.01$, $t/\Gamma_0 =0.8$, $\alpha=0.15$, $U=0$, $\gamma = 1$, and
$p=0$. The insets in (a) and (b) show respectively the thermopower
$S$ and figure of merit $ZT$ for $q=1$.}
\end{figure}

Let us begin with the case of vanishing Coulomb interaction,
$U=0$. In Fig.\ref{Fig:1} the thermoelectric quantities are
plotted as a function of the dots' energy levels $\varepsilon_0$ for
$t/\Gamma_0 =0.8$ and for indicated values of the parameter $q$.
The latter parameter effectively describes strength of the
indirect (via the leads) coupling of the dots. Such a coupling
contributes (like direct inter-dot hoping term) to the formation
of bonding and antibonding states. As a result, a broad peak
corresponding to the bonding state and a narrow one corresponding
to the antibonding state emerge in the density of
states.~\cite{trocha07} For $q$ close to 1, the conductance peak
associated with the antibonding state reveals the antiresonance
character with a characteristic Fano line shape, whereas the
conductance peak associated with the bonding state is relatively
broad and roughly Lorentzian. For smaller (but nonzero) values of
the parameter $q$, the antiresonance is suppressed and one
observes two peaks of different widths. Finally, for $q=0$ two
peaks of equal widths emerge in the linear conductance. Similar
line shape and $q$ dependence is observed in the electronic
contribution to the heat conductance, as shown in
Fig.\ref{Fig:1}(c). Positions of the peaks in thermal conductance
correspond  well to those in the electric conductance, similarly
as in the case of a single dot.~\cite{swirkowicz}

The thermopower $S$, shown in Fig.\ref{Fig:1}(a), changes sign
when $\varepsilon_0$ corresponds to one of the relevant resonances.
This is a consequence of the compensation of charge current due to
electrons by that due to holes. As a result, there is no net
charge current and no voltage drop, and consequently the
thermopower vanishes. When the energy level is located below the
resonance, the thermopower is negative because the majority
carriers are holes. In turn, when the energy level is above the
resonance the main carriers are electrons and thus the thermopower
is positive. Note that $S$ in Fig.1(a) is measured in the units of
$k_B/e$, with $e$ denoting electron charge ($e<0$). However, when
$q$ is close to 1, one finds two more points where the thermopower
changes sign. One of them is situated near the Fano resonance,
where the conductance vanishes due to destructive interference,
whereas the other one is located in the valley between the two
resonances. For other values of $q$ only the point in the valley
between the resonances is present, as no complete destructive
interference occurs for smaller $q$. The largest increase of the
thermopower appears in the vicinity of the antibonding resonance
and for $q=1$. This shows that the quantum interference has a huge
impact on the thermoelectric phenomena. This is also clearly
visible in the figure of merit $ZT$, shown in Fig.\ref{Fig:1}(b),
which is considerably enhanced in the vicinity of the Fano
antiresonance, where $ZT$ is close to 1. Outside this region $ZT$
is significantly suppressed, even near the bonding energy level.
When  $q$ decreases, the figure of merit $ZT$ (which reaches
almost unity for $q=1$ near the Fano resonance) diminishes as
well. Thus, the interference effects play a crucial role in
thermoelectric efficiency of DQD systems considered here. However,
in this coupling regime ($k_BT\ll \Gamma$) the Wiedemann-Franz law
is not violated much as the Lorenz number (not shown) remains
close to unity.

\begin{figure}
\includegraphics[width=0.34\textwidth,angle=0]{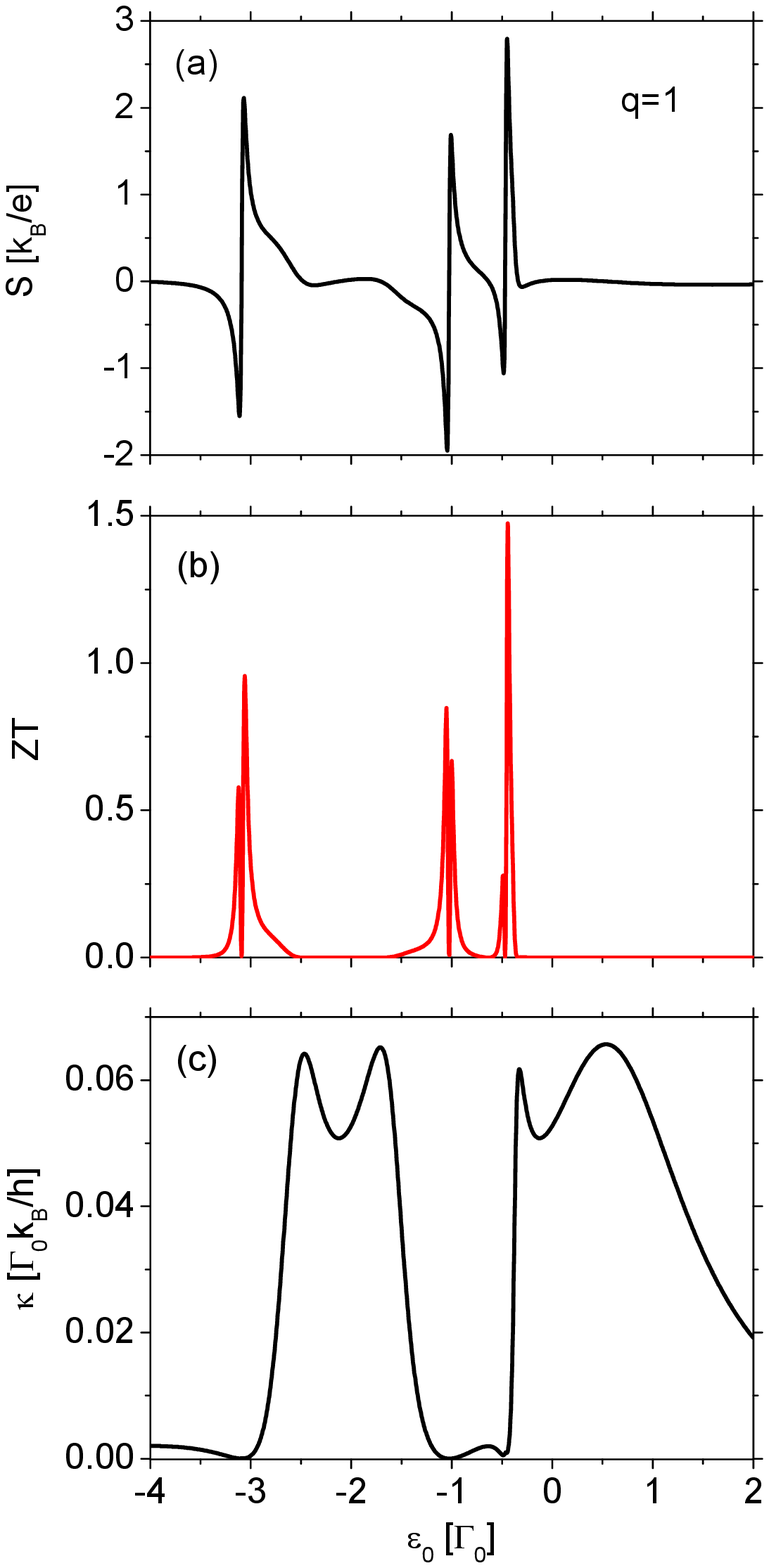}
\caption{\label{Fig:2}(color online) Thermopower $S$ (a), figure
of merit $ZT$ (b), and the thermal conductance (c), calculated  as
a function of the dots' levels energy for  $U/\Gamma_0 =2$  and
$q=1$. The other parameters as in Fig.\ref{Fig:1}.}
\end{figure}

\begin{figure}
\includegraphics[width=0.3\textwidth,angle=0]{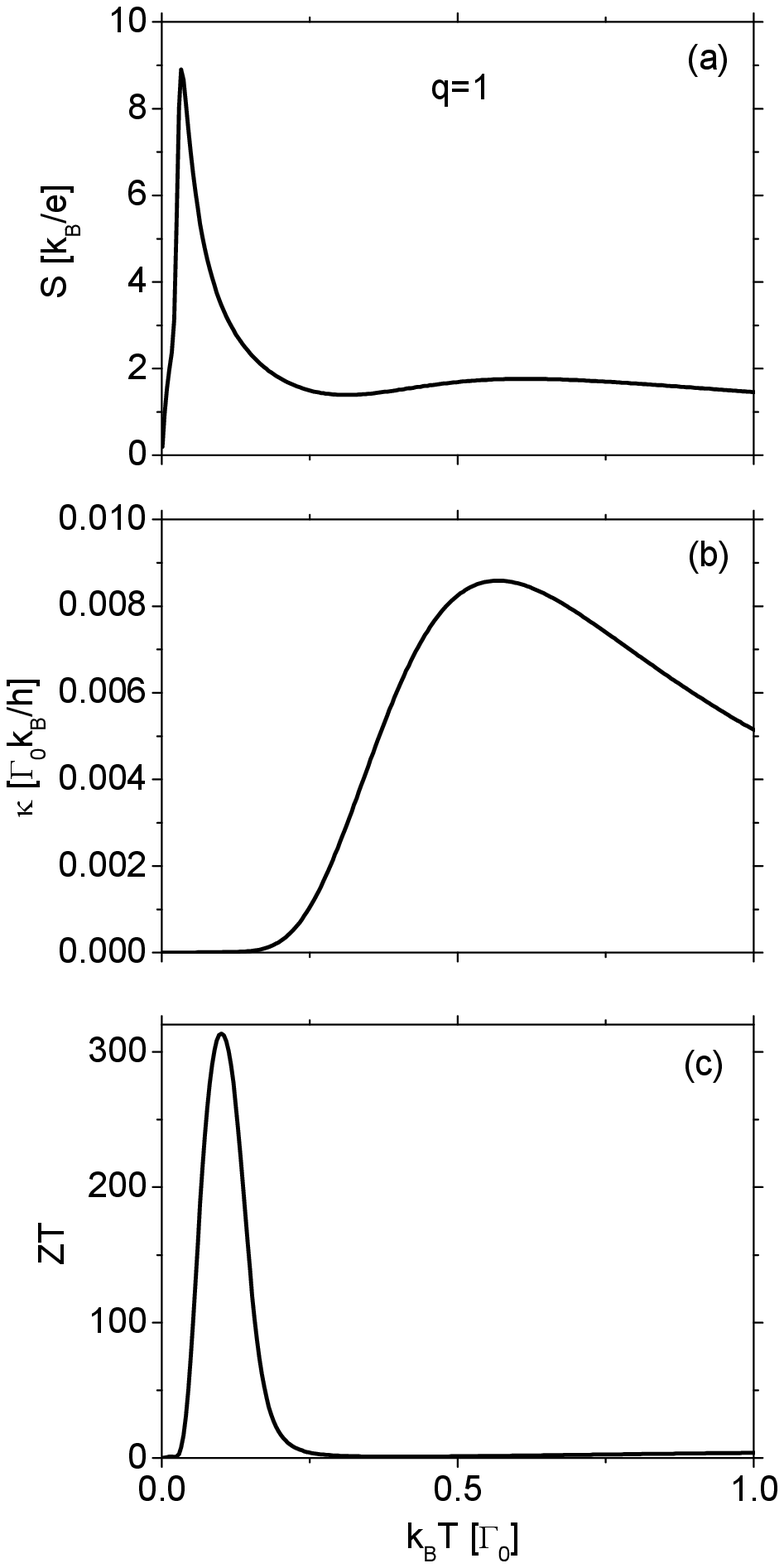}
\caption{\label{Fig:3}(color online) Thermoelectric coefficients:
(a) thermopower, (b) thermal conductance, (c) figure of merit,
calculated as a function of temperature for the position of the
dots' levels in the vicinity of the Fano resonance,
$\varepsilon_0/\Gamma_0 =-1.15$, and for $q=1$, $U=0$, and
$\gamma=0.01$. The other parameters as in Fig.\ref{Fig:1}.}
\end{figure}

Consider now the impact of Coulomb repulsion on the thermoelectric
properties in the presence of Fano resonance, see Fig.\ref{Fig:2}.
First, the Coulomb interaction leads to splitting of the
double-peak structure (present for $U=0$) in both the charge and
thermal conductance,  and characteristic Coulomb gaps
occur\cite{trocha07} (see Fig.\ref{Fig:2}(c) for the thermal
conductance).  This doubling of the resonances also leads to a
richer structure of the thermopower. For a finite $U$ the
thermopower changes sign ten times, see Fig.\ref{Fig:2}(a). Four
points of zero $S$ are associated with the four resonances located
roughly at  $\pm t$ and $\pm t-U$. Next two such points are
located near the Fano peak and its Coulomb counterpart, where the
conductance disappears (so does the thermopower), whereas two
other points are situated in the valleys between the narrow and
broad maxima of the conductance. In the latter case the
thermopower vanishes due to weighted-symmetry in location of the
bonding and antibonding states with respect to the Fermi level.
The current due to electrons tuneling through the bonding state is
compensated then by the current due to holes tunneling through the
antibonding level. This is a {\emph 'local'} bipolar effect. The
thermopower disappears also in the symmetry point
$\varepsilon=-U/2$, as it has been explained in
Ref.[\onlinecite{swirkowicz}]. However, there is one more point
where the thermopower changes sign, namely this happens at the
energy where a small maximum appears in the Coulomb gap (this
feature is due to a remnant of the conductance maximum
corresponding to the antibonding state). Correspondingly, more
peaks occur in the figure of merit, see Fig.\ref{Fig:2}(b). From
Figs 1 and 2 also follows that the Coulomb interactions can
increase the magnitude of $ZT$, which now exceeds 1. However, $ZT$
is considerably enhanced only near the Fano antiresonances (as in
the case of $U=0$) and also in the middle of the Coulomb gap. The
latter is due to a global bipolar effect.

\begin{figure}
\includegraphics[width=0.34\textwidth,angle=0]{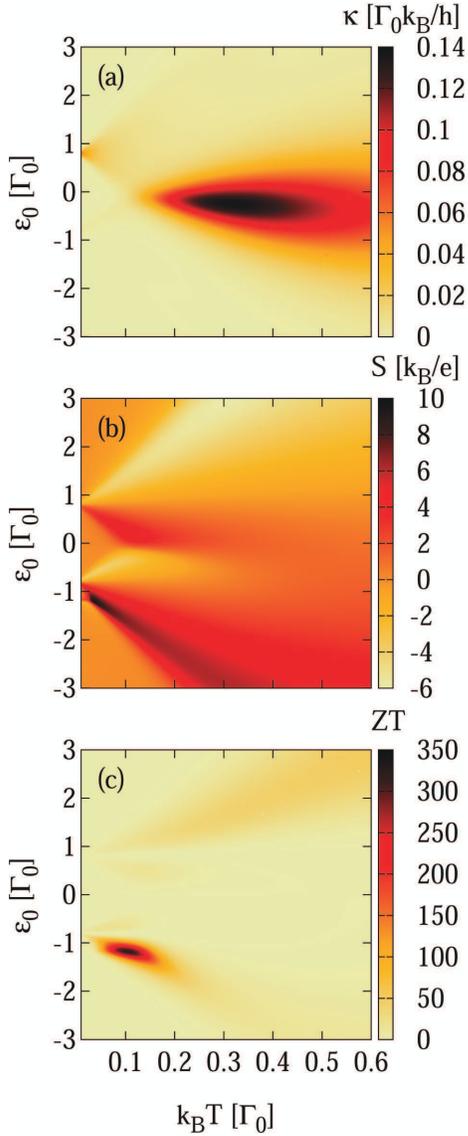}
\caption{\label{Fig:4} Thermal conductance  (a), thermopower (b),
and the figure of merit (c), as a function of temperature and
dots' levels energy for $q=1$, $U=0$, $\gamma=0.01$, and the other
parameters as in Fig.\ref{Fig:1}.}
\end{figure}

\begin{figure}
\includegraphics[width=0.34\textwidth,angle=0]{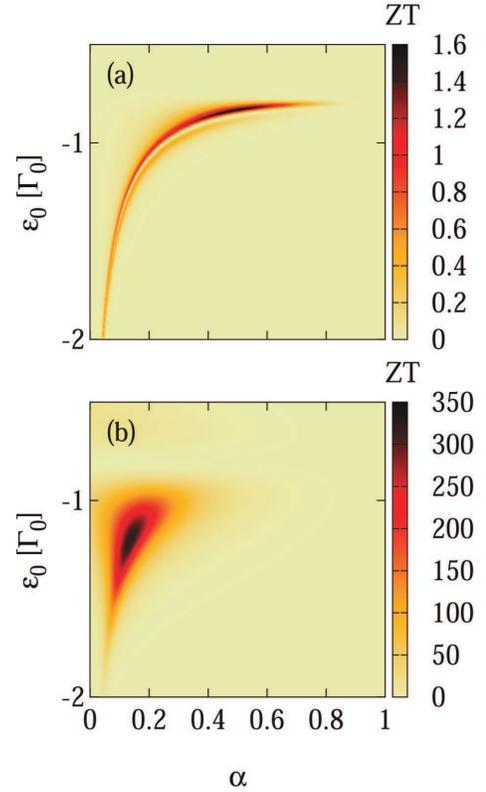}
\caption{\label{Fig:5} Figure of merit as a function of the
asymmetry parameter $\alpha$  and the dots' levels energy,
calculated for $k_BT/\Gamma_0=0.01$ and $\gamma=1$ (a), and
$k_BT/\Gamma_0=0.1$ and $\gamma=0.01$ (b). Apart from this, $q=1$,
$U=0$, and the other parameters as in Fig.\ref{Fig:1}.}
\end{figure}

Above we have analyzed the range of $k_BT/\Gamma\ll 1$. The thermal
efficiency, however,  strongly depends on the ratio of thermal
energy and coupling strength, and for $k_BT/\Gamma > 1$ one might
expect an increase of the thermal efficiency.~\cite{murphy} On the
other side, however, higher temperature suppresses the
interference effects responsible for the Fano antiresonance. In
Fig.~\ref{Fig:3} we show the temperature dependence of the
thermoelectric coefficients for the level position near the
antibonding state and vanishing Coulomb interactions. The  results
clearly show that the thermoelectric effects are optimized for
$k_BT/\Gamma\approx 10$, or equivalently $k_BT/\Gamma_0\approx
10\gamma$. To show this more explicitly, in Fig.~\ref{Fig:4} we
present the temperature and level position dependence of the
thermoelectric coefficients. Figure~\ref{Fig:4}(c) clearly shows
that the largest enhancement of the thermoelectric efficiency
occurs in the vicinity of the antibonding level. Moreover, as the
temperature grows, a strong increase of the thermal conductance is
observed in the valley between the two low-temperature maxima of
the conductance, see Fig.~\ref{Fig:4}(a). Such a peak is absent in
the electric conductance and also does not appear in thermal conductance
in the low temperature regime. When the temperature is relatively high, the
Fermi-Dirac distribution around the Fermi level becomes smeared,
and electrons of higher energies are involved in transport. As the
energy of tunneling electrons does not influence the electric
conductance, it plays a crucial role in the thermal conductance.
As already mentioned above, in the symmetry point,
$\varepsilon_0=-U/2$, the charge current due to electrons is
compensated by current due to holes. However, both electrons and holes 
flow in the same direction, so the energy carried by
both types of carriers is not compensated, in contrast to the
charge. As a result, an additional peak centered at
$\varepsilon_0=-U/2$ appears in the thermal conductance for a
sufficiently high temperature.

The thermoelectric efficiency of the system under consideration
can be further optimized by tuning asymmetry (parameter $\alpha$)
in the coupling of a given lead to the two dots. To show this, in
Fig.~\ref{Fig:5} we present the $\alpha$ and level position
dependence of the figure of merit for two distinct coupling
regimes: $k_BT/\Gamma\ll 1$ (part (a)) and  $k_BT/\Gamma
> 1$ (part (b)). The thermoelectric efficiency in these
two transport regimes is optimized for different ranges of the
parameter $\alpha$. n the low temperature regime ($k_BT/\Gamma\ll 1$), $ZT$ is optimized for
intermediate values of the asymmetry parameter, roughly for
$\alpha\in\langle 0.3,0.6\rangle$. In turn  for $k_BT/\Gamma
> 1$, the figure of merit achieves the highest values for
relatively high asymmetry, $\alpha\in\langle 0.1,0.2\rangle$. It
is worth noting that in the former case the figure of merit $ZT$
can exceed $1$ for properly chosen asymmetry parameter $\alpha$
despite of relatively strong dot-lead coupling, while in the
latter case $ZT$ can reach extremely large values exceeding 300.

%1comment
The strong dependence of  $ZT$ on the asymmetry parameter $\alpha$, displayed in Fig.~\ref{Fig:5} (a), is due to the interference effects of electron waves passing through different paths in the system. More specifically, it originates from the $\alpha$ dependence of the position of the point where the electron conductance (as well as the thermal conductance) reaches almost zero due to the destructive quantum interference. At this point (and in its neighborhood) the thermoelectric effects are maximized, which is reflected by enhanced values of the figure of merit $ZT$ in Fig.~\ref{Fig:5} (a). In turn, in the high temperature regime, Fig.~\ref{Fig:5} (b), the Fermi-Dirac distribution around the Fermi level becomes smeared and since the low and high energy electrons contribute differently to heat current, the thermal conductance peaks do not coincide now with those in the electric conductance  (contrary to  the low temperature regime where they do). Thus, behavior of the thermal conductance is now more complex. In the vicinity of the antibonding state,  the thermal conductance is strongly suppressed achieving local minimum. This region of suppressed thermal conductance appears in the 'tail' of the electric conductance peak, where $G$ is finite. Moreover, position of the minimum in thermal conductance depends on the asymmetry parameter $\alpha$, while position of the peak in electric conductance is rather unchanged. This behavior leads to a large enhancement of $ZT$ seen in Fig.~\ref{Fig:5}(b) for $0.1 < \alpha <0.2$. When $\alpha$ increases above $0.2$, the intensity of the peak in electric conductance corresponding to the antibonding state becomes suppressed, and this leads to suppression of $ZT$ for $\alpha > 0.2$. On the other hand the suppression of $ZT$ for $\alpha<0.1$ is associated with a strong dependence of the minimum position in the thermal conductance on the parameter $\alpha$ for $\alpha\in (0,0.2)$.

\begin{figure}
\includegraphics[width=0.4\textwidth,angle=0]{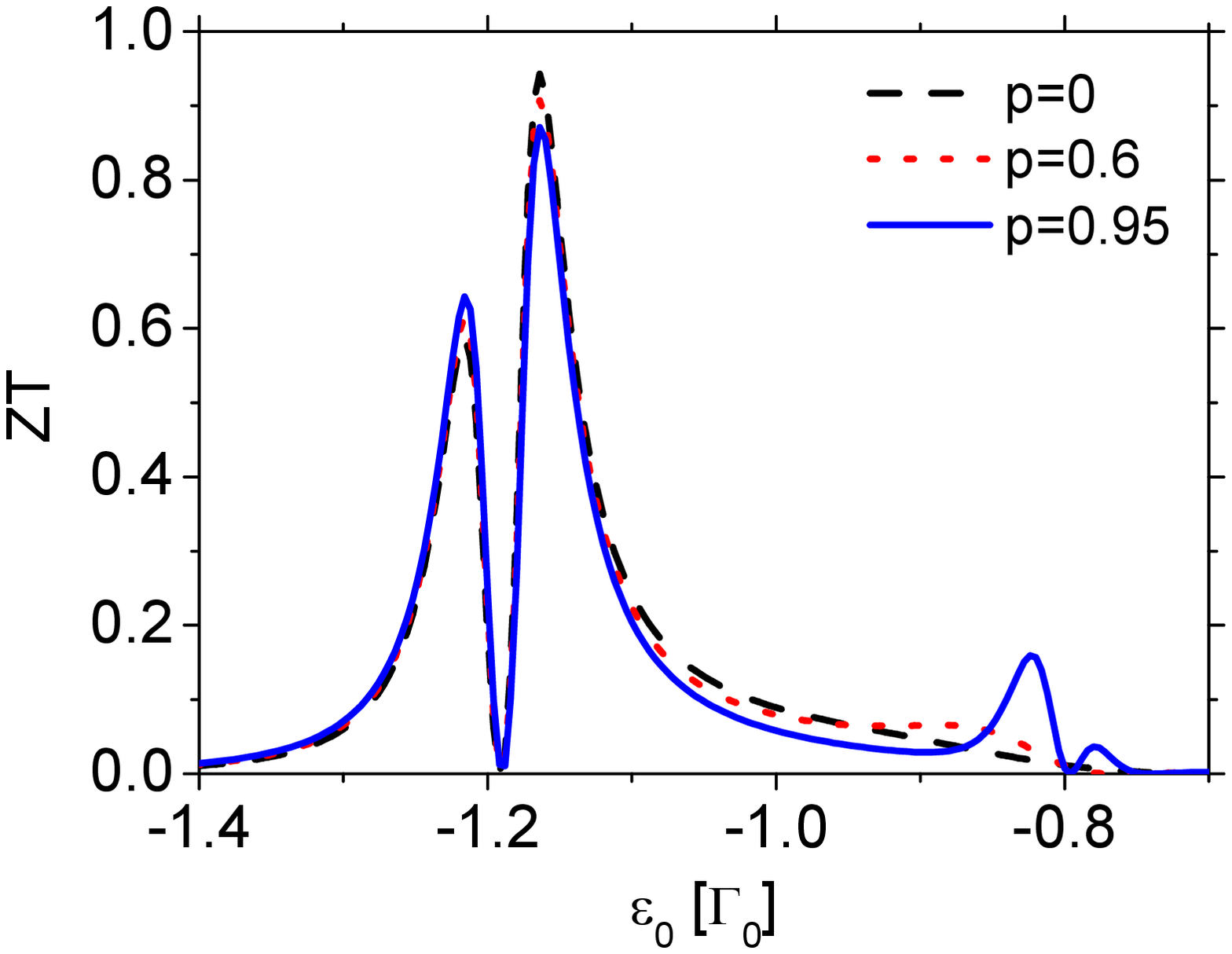}
\caption{\label{Fig:6}(color online) Thermoelectric figure of
merit $ZT$ as a function of the dots' levels energy, calculated
for indicated values of the leads' polarization $p$ in the
parallel magnetic configuration, and for $q=1$, $U=0$,
$k_BT/\Gamma_0=0.01$ and $\gamma=1$. The other parameters as in
Fig.\ref{Fig:1}. }
\end{figure}

\subsubsection{Magnetic leads: $p>0$}\label{Sec:3aa}

Now we consider the situation when both electrodes are
ferromagnetic, with the corresponding spin polarization factor
$p$. In the following only collinear, i.e. parallel and
antiparallel, configurations will be analyzed.

\begin{figure}
\includegraphics[width=0.33\textwidth,angle=0]{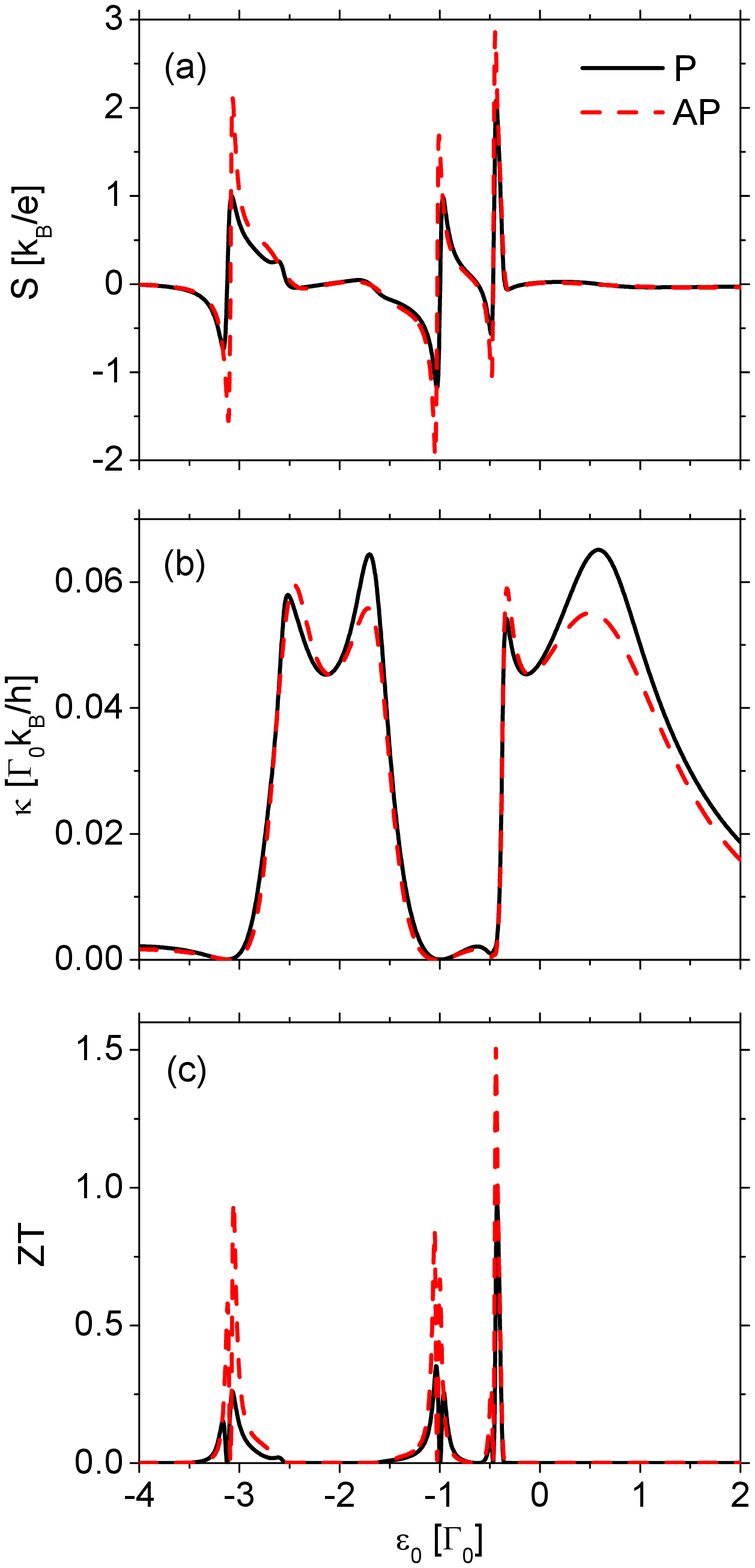}
\caption{\label{Fig:7}(color online) Thermoelectric coefficients:
(a) thermopower, (b) thermal conductance, (c) the figure of merit
as a function of the dots' levels energy, calculated for parallel
(P) and antiparallel (AP) magnetic configurations, and for
$p=0.4$,  and $U/\Gamma_0 =2$. The other parameters as in
Fig.\ref{Fig:6}.}
\end{figure}

In the coupling regime $k_BT/\Gamma\ll 1$, the Seebeck coefficient
and figure of merit are only weakly dependent on the leads'
polarization. As it has been mentioned above, $ZT$ is enhanced
only in the vicinity of the Fano resonance. For $U=0$, one can
notice the double narrow peak structure in $ZT$ near the Fano
resonance, see Fig.\ref{Fig:6}, where $ZT$ is shown for three
different values of the spin polarization factor $p$ (including
for comparison also the case of $p=0$). Between the peaks $ZT$
reaches zero due to vanishing thermopower (as already explained
above). However, an additional double peak feature in $ZT$ appears
for sufficiently large $p$, with the intensities (widths)
increasing (decreasing) with increasing polarization. This feature
is absent for small values of $p$ (also for $p=0$, see the inset
to Fig.1(b)). The appearance of the second double peak structure
can be accounted for as follows: Since the coupling between the
dots and external (magnetic) leads is spin-dependent for $p\ne 0$,
the level widths of the bonding and antibonding states are spin
dependent, too. This results in narrowing (broadening) of the
conductance peak corresponding to the spin down (spin up)
carriers. Consequently, the characteristic zero-conductance Fano
point is achieved first in the spin-down channel when going
towards larger negative values of $\varepsilon_0$. This, in turn,
results in well resolved sharp features in the Seebeck coefficient
- one coming from spin-up contribution and another one (smaller)
from the spin-down contribution. As a result, additional
double-peak structure appears in the thermal conductivity (not
shown) and also in the figure of merit, as in Fig.\ref{Fig:6}. The
role of finite $U$ is similar to that described for $p=0$.

\begin{figure}[t]
\includegraphics[width=0.33\textwidth,angle=0]{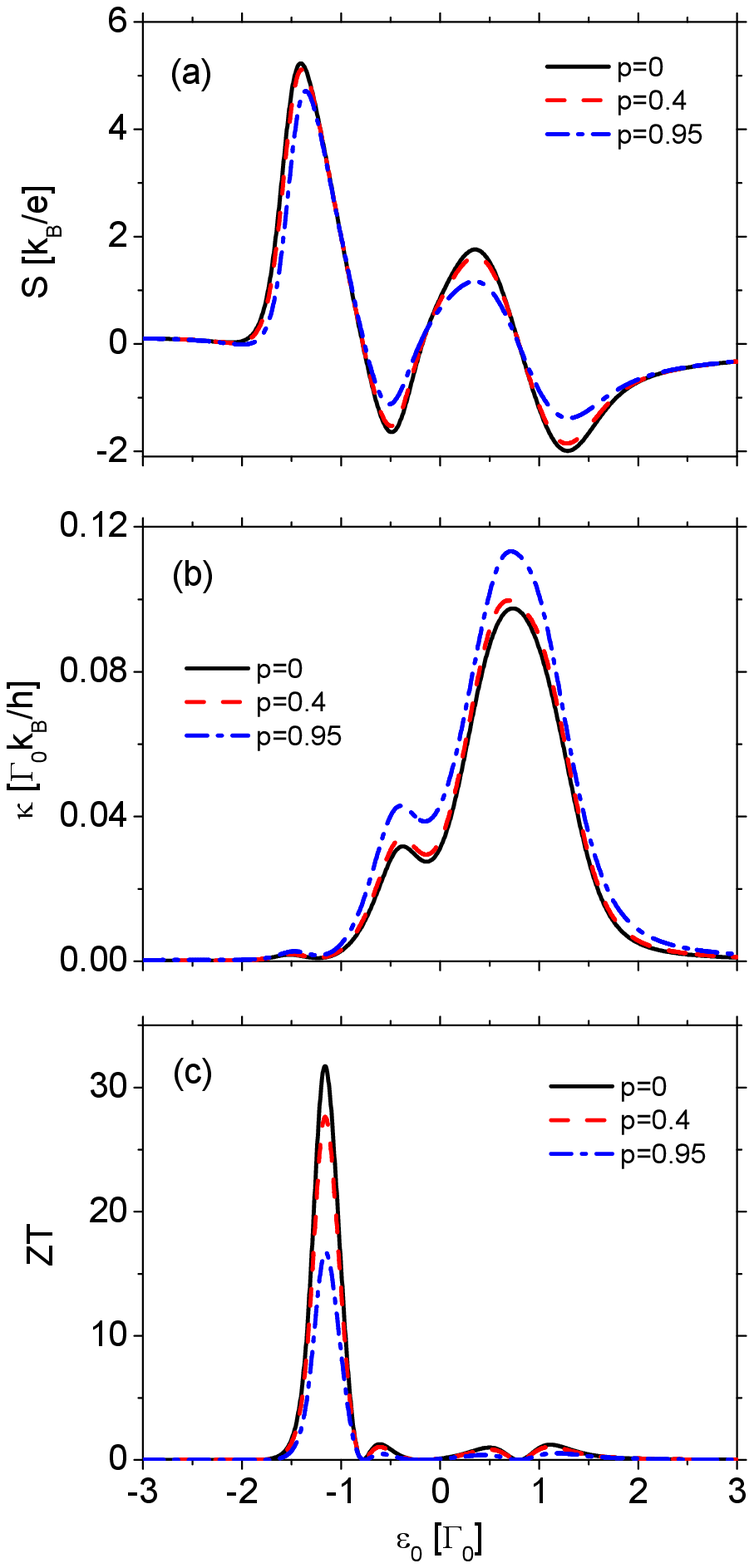}
\caption{\label{Fig:8}(color online) Thermoelectric coefficients:
(a) thermopower, (b) thermal conductance, (c) the figure of merit
as a function of the dots' levels energy, calculated for indicated
values of the leads' polarization $p$ in the parallel magnetic
configuration, and for $U/\Gamma_0 =0$, $q=1$,
$k_BT/\Gamma_0=0.1$, and $\gamma=0.1$ ($k_BT/\Gamma=1$).  The
other parameters as in Fig.\ref{Fig:1}. }
\end{figure}

The thermoelectric coefficients depend now on magnetic
configuration of the system. In Fig.\ref{Fig:7} the thermopower,
thermal conductance and figure of merit $ZT$ are displayed for the
parallel and antiparallel magnetic configurations and for a finite
$U$. It is worth noting that both thermopower and figure of merit
are larger in the antiparallel configuration, whereas the thermal
conductance is larger in the parallel configuration. The
difference between $S$ (and also $ZT$) in the parallel and
antiparallel configurations is clearly visible  at the
corresponding peaks, where it can be quite significant. In turn,
the difference between heat conductances in both configurations is
rather small and well resolved only in the vicinity of the
relevant resonances. Anyway, Fig.\ref{Fig:7} reveals the
possibility of constructing a \emph{heat spin valve}, which would
be a heat analog of the usual current spin valve.
%2comment
The suppression of thermal conductance in the antiparallel magnetic configuration is of similar origin as the suppression of linear electric conductance. The later was accounted for by Julliere in terms of the two current model~\cite{Julliere}, and is due to different densities of states at the Fermi level in both configurations for electrons of a given spin orientation. More specifically, in the parallel configuration there is one spin channel (corresponding to high density of states in the source and drain electrodes) with high electric (and also heat) conductance  and one with low conductance (corresponding to low density of states in the source and drain electrodes). In turn, in the antiparallel configuration both conduction channels have reduced conductance as now each of them corresponds to high density of states in one electrode and low density of states in the second electrode.

\begin{figure}
\includegraphics[width=0.36\textwidth,angle=0]{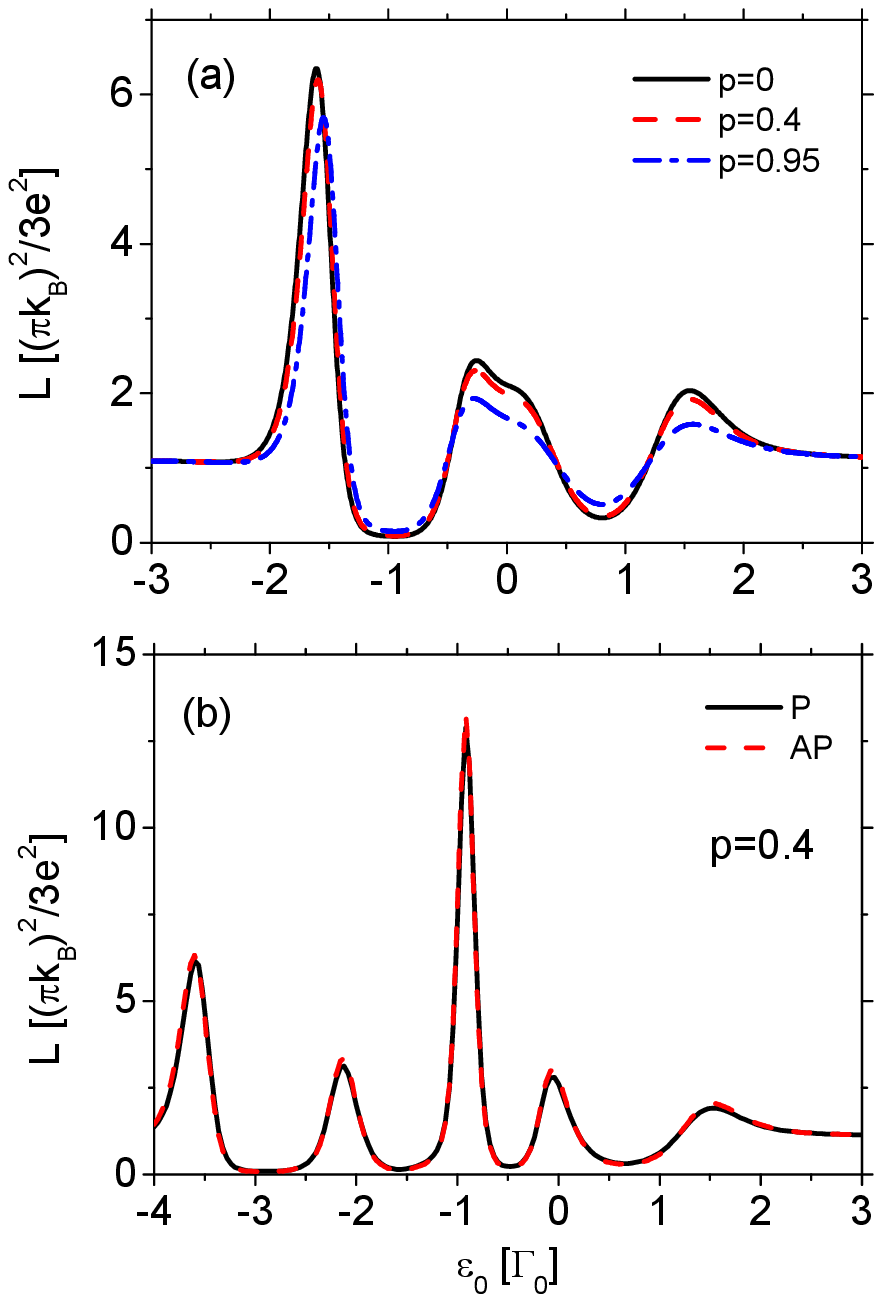}
\caption{\label{Fig:9}(color online) Lorentz ratio as a function
of the dots' levels energy calculated for indicated values of the
leads' polarization $p$. Part (a) corresponds to parallel
configuration and $U=0$, while part (b) to $U/\Gamma_0=2$ and for
parallel (P) and antiparallel (AP) magnetic configurations. The
other parameter as in Fig.\ref{Fig:8}.}
\end{figure}

Now let us discuss briefly the temperature regime
$k_BT/\Gamma_0\approx\gamma$ ($k_BT/\Gamma\approx 1$). In
Fig.~\ref{Fig:8} the thermoelectric coefficients are displayed for
indicated values of the leads' polarization, no Coulomb interaction, 
and for parallel magnetic configuration. One can
notice, that the thermal conductance increases with increasing
leads' polarization $p$. In turn, the magnitudes of the
thermopower as well as of  the figure of merit decrease with
increasing $p$. Similar behavior has been also reported for a
single quantum dot attached to ferromagnetic
leads.~\cite{swirkowicz}

\begin{figure}
\includegraphics[width=0.33\textwidth,angle=0]{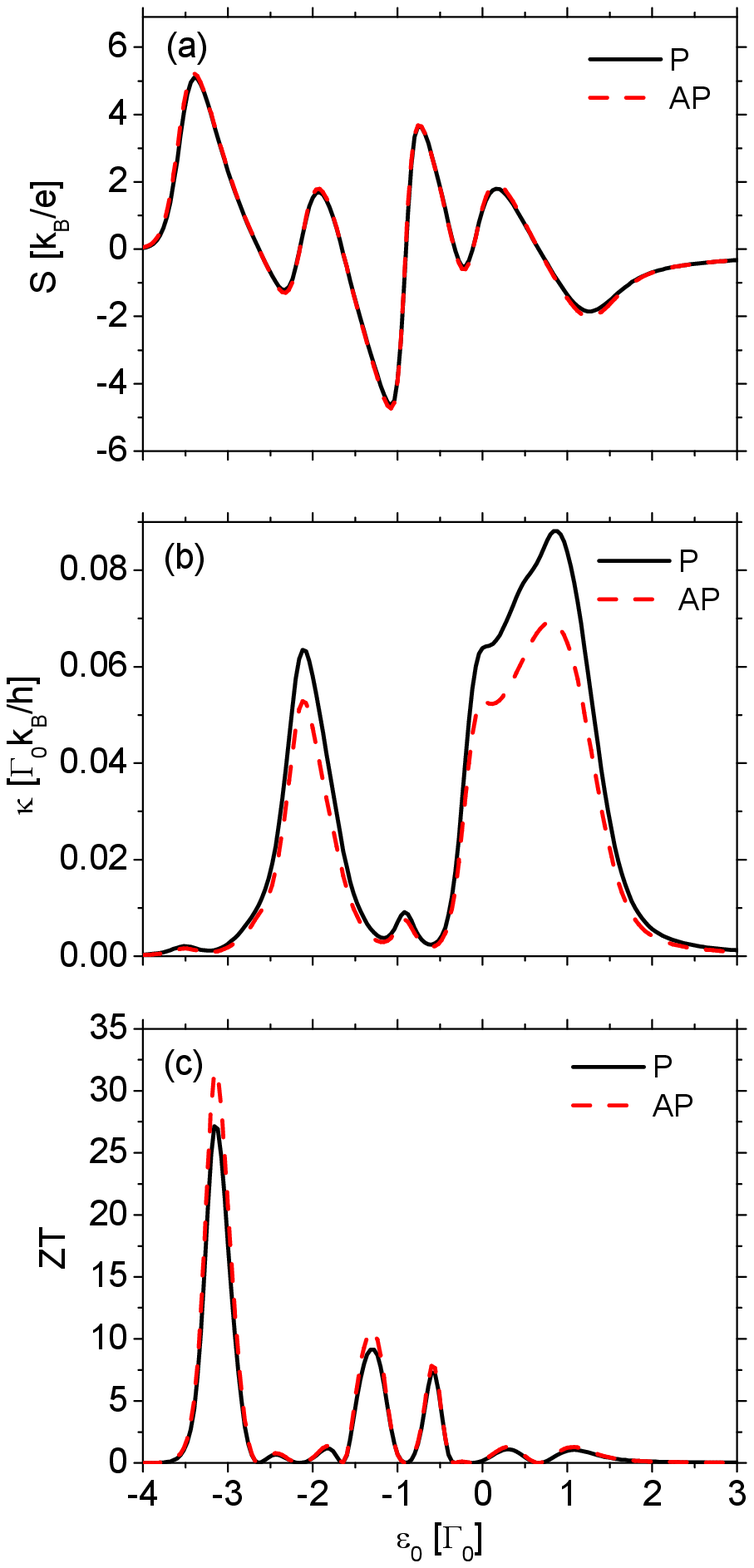}
\caption{\label{Fig:10}(color online) Thermoelectric coefficients:
(a) thermopower, (b) thermal conductance, (c) the figure of merit,
as a function of the position of the dots' energy levels
calculated for parallel (P) and for antiparallel (AP) magnetic
configurations, and for $p=0.4$. The other parameters as in
Fig.\ref{Fig:8}.}
\end{figure}

Although the quantum interference effects become {\it smeared out}
by increasing temperature, the largest changes in the thermopower
 for $k_BT/\Gamma_0\approx 1$ still occur in the vicinity of the antibonding level, where the
conductance is strongly suppressed for $k_BT/\Gamma\ll 1$. Therefore, the
figure of merit $ZT$ achieves there relatively high values (above
30 in Fig.8(c)). All this leads to violation of the
Wiedemann-Franz law, which can be measured by the Lorenz ratio,
$L=\kappa/GT$, as shown explicitly in  Fig.~\ref{Fig:9}, where the
Lorenz ratio is displayed for indicated values of the leads'
polarization $p$ and for interacting ($U\ne 0$) and
non-interacting ($U= 0$) cases. One can note that the Coulomb
interactions lead  to larger deviations from the Wiedemann-Franz
law, especially in the vicinity of the symmetry point
$\varepsilon_0=-U/2$~[Fig.~\ref{Fig:9}(b)].

Similarly as for the temperature range studied before, thermoelectric
coefficients depend on magnetic configuration, too. By changing
magnetic configuration of the system from parallel to antiparallel
one, the thermoelectric efficiency increases and is comparable to
that obtained for $p=0$. This is because in the antiparallel
magnetic configuration both electric and thermal conductance
decrease (see Fig.~\ref{Fig:10}(b) for the thermal conductance).
However, the latter suffers a larger drop and thus the figure of
merit increases.

\begin{figure*}
\includegraphics[width=0.8\textwidth,angle=0]{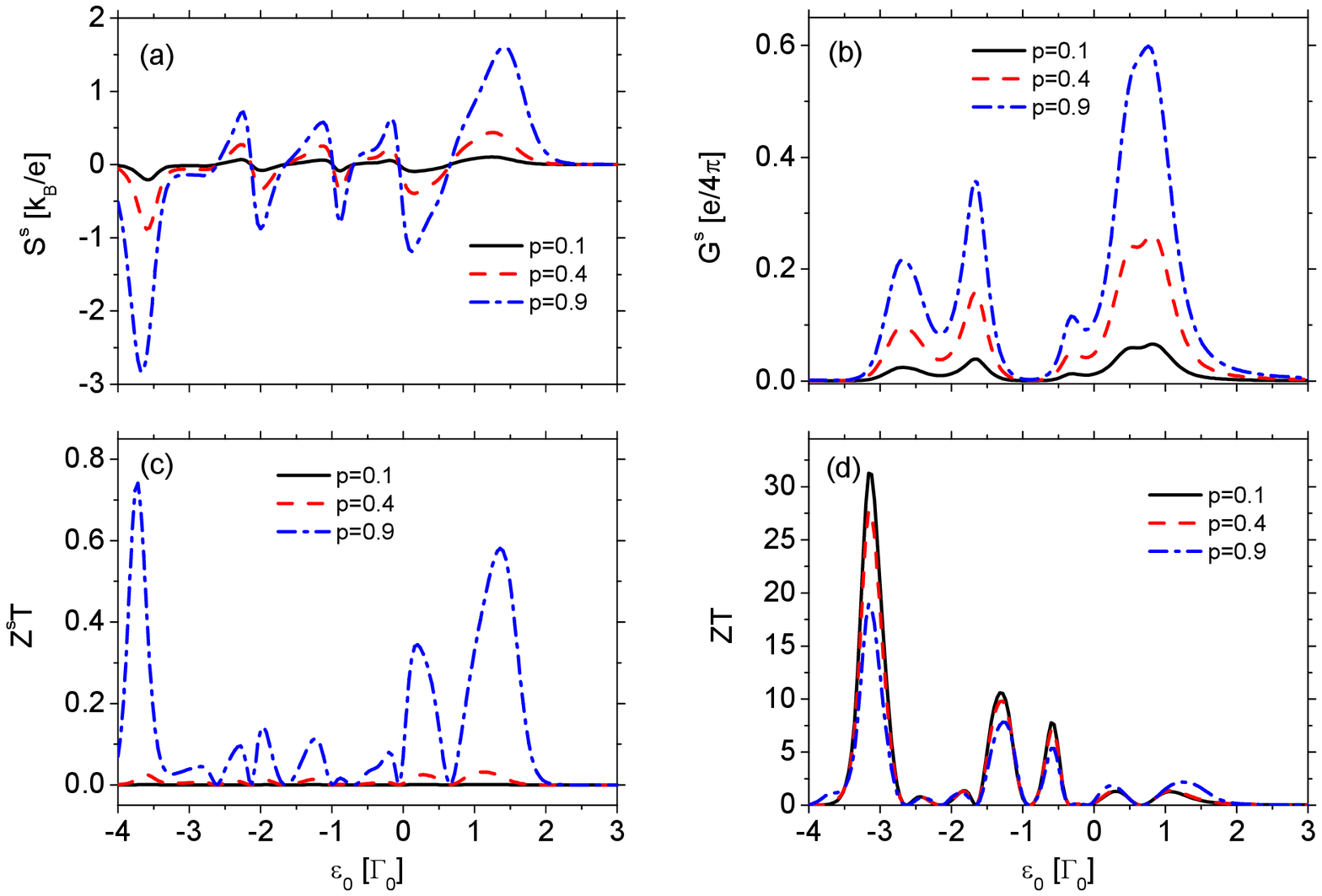}
\caption{\label{Fig:11} Spin thermoelectric coefficients: (a) spin
thermopower, (b) spin conductance, (c) $Z^sT$, (d) $ZT$,
calculated as a function of the dots' levels energy for indicated
values of the leads polarization $p$ in the parallel magnetic
configuration. The other parameters: $k_BT/\Gamma_0 =0.1$,
$\gamma=0.1$, $t/\Gamma_0=0.8$, $\alpha=0.15$, $U/\Gamma_0=2$,
$q=1$. }
\end{figure*}

\subsection{Spin bias and spin thermoelectric
effects}\label{Sec:3b}

Assume now that spin accumulation in the external leads becomes
relevant, which may happen when spin relaxation time in the leads
is sufficiently long. In such a case we have to take into account
spin splitting of the electro-chemical level in both leads.
Accordingly, the bias is then spin dependent and the difference in
chemical potentials, $\Delta\mu_\sigma$, in the spin channel
$\sigma$ can be written as (see section 2 for details);
$\Delta\mu_\sigma \equiv eV_\sigma =e(V+\hat{\sigma} V^s)$. Of
course, the spin bias $V^s$ vanishes in the absence of spin
accumulation.

\begin{figure*}
\includegraphics[width=0.8\textwidth,angle=0]{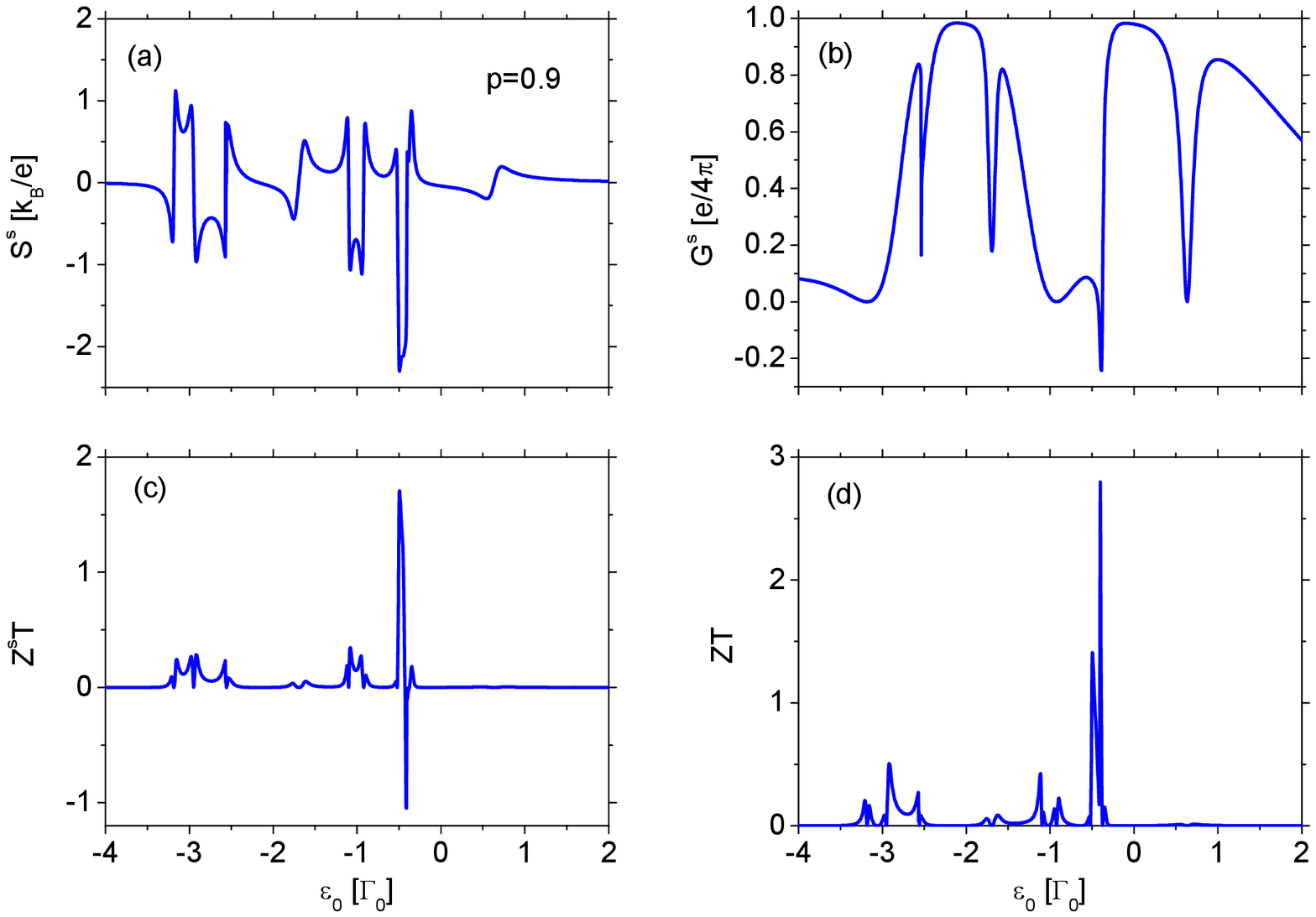}
\caption{\label{Fig:12} Spin thermoelectric coefficients: (a) spin
thermopower, (b) spin conductance, (c) $Z^sT$, (d) $ZT$,
calculated as a function of the position of the dot levels for
indicated values of the leads polarization $p$ in the parallel
magnetic configuration. The other parameters:
$k_BT/\Gamma_0=0.01$, $\gamma=1$. The other parameters as in
Fig.11. }
\end{figure*}

As we have already mentioned in the introduction,  thermally
induced spin voltage has been observed recently in ferromagnetic
metallic slabs.~\cite{uchida} A consequence of this is a spin
analog of the Seebeck effect, so-called spin Seebeck effect (or
spin thermopower). In the following we present some numerical
results for the spin thermoelectric effects in the system under
consideration, especially for the spin thermopower and spin analog
of the figure of merit. The latter quantity is defined as
$Z^sT=(2e/\hbar) G^s{S^{s}}^2T/\kappa$

In Fig.~\ref{Fig:11} we present the spin thermoelectric
coefficients in the case of $k_BT/\Gamma = 1$. Since the basic
features of the thermopower $S$ are similar to those obtained and
discussed  earlier, we focus only on the spin thermopower $S^{s}$,
see Fig.11(a). The spin thermopower depends on the leads'
polarization more strongly than the thermopower $S$. Note, $S^{s}$
vanishes in the limit of $p=0$  and $|S^s|$ grows up with
increasing $p$. Comparing Fig.~\ref{Fig:11} and Fig.~\ref{Fig:8}
one finds that the sign of the spin thermopower $S^s$ is opposite
to the sign of the charge thermopower $S$. This is because the
voltage drop induced in the spin minority channel is larger than
that induced in the spin majority channel. Generally, the spin
thermopower vanishes at the same level positions as the charge
thermopower does. Similar behavior refers to the spin conductance
shown in Fig.~\ref{Fig:11}(b). It vanishes for $p=0$ and grows
with increasing $p$, being positive in Fig.11(b).  The spin analog
of the figure of merit, $Z^sT$, is shown in Fig.~\ref{Fig:11}(c).
It obviously grows up with increasing $p$ as a result of the
increase of both $|S^s|$ and $G^s$ and a rather weak  dependence
of the thermal conductance on the leads polarization.  The
corresponding figure of merit $ZT$ is shown in
Fig.~\ref{Fig:11}(d). $ZT$ calculated on the condition of
vanishing currents in both spin channels depends on the magnetic
polarization of the leads in a more complex way than $ZT$ obtained
on the condition of $J=0$. Roughly speaking, it grows up (drops)
with increasing $p$ for $\varepsilon_0>0$
($\varepsilon_0<0$). The suppression of $ZT$ corresponds to the
regions where the thermopower $S$ is almost independent of the
spin polarization of the leads.

Qualitatively similar behavior of the spin thermoelectric
coefficients can be observed also in the low temperature regime,
$k_BT/\Gamma\ll 1$, as shown in  Fig.~\ref{Fig:12} for a
relatively high polarization factor, $p=0.9$. The maximum value of
spin figure of merit $Z^sT$ is now slightly higher than that in
Fig.11, contrary to $ZT$ which now is significantly smaller than
in the case shown in Fig.11. Apart from this, one feature deserves
mentioning here, i.e. the appearance of a negative spin
conductance, which appears in some regions of the dots' energy
levels. This, in turn, leads to negative values of the spin figure
of merit $Z^sT$, see Fig.~\ref{Fig:12}(c). Moreover, the spin
conductance is now enhanced in comparison to that shown in
Fig.11(b).

\section{Summary}\label{Sec:4}

We have analyzed thermoelectric effects, like thermopower,
electronic contribution to heat conductance, and thermoelectric
efficiency in a system of two coupled quantum dots. The key point
of the analysis was the role of interference effects (especially
of the Fano antiresonances). We have shown that the interference
effects can significantly enhance the thermoelectric efficiency,
especially for specific thermal energy ranges. The thermoelectric
efficiency can be additionally enhanced by Coulomb interactions.
Thus, the thermoelectric efficiency can be controlled by
temperature, Coulomb interactions and quantum interference
effects. Moreover, we have also shown that the thermoelectric
efficiency can be additionally  controlled by asymmetry of the
coupling of a given electrode to the two dots.
%4comment
Some of the parameters, especially coupling between the two dots and between the dots and leads can be tuned by external gate voltages. It is more difficult to control Coulomb parameter $U$, although it can be  also tuned by  changing size of the dots via additional gates.

We have also considered spin thermoelectric phenomena, in
particular the spin analog of the thermopower -- the so-called
spin thermopower or spin Seebeck effect. The latter effect occurs
as a result of spin asymmetry of the two spin channels. From the
experimental point of view, the spin thermopower may be observed
when the spin relaxation time in the external leads is
sufficiently long, so the spin accumulation may build up. We have
also analyzed the corresponding spin figure of merit.

As we have already mentioned above, the interference effects (like
Fano resonance) can significantly enhance the thermoelectric
efficiency of a device. The results presented include only
electronic contribution to the heat current. However, one should
bear in mind that there is also an additional phonon term in the
heat conductivity, and this contribution to heat transport may
reduce the thermoelectric efficiency presented above, especially
at higher temperatures.~\cite{Snyder,Stojanovic} At low
temperatures, however, this contribution is small and therefore
the results reported in this paper present a reasonable and
satisfactory description.
%1comment
The formula for thermoelectric figure of merit, including the phonon contribution, can be written as $ZT=ZT_e/(1+\kappa_{ph}/\kappa_e)$, where $\kappa_{ph}$  is the phonon contribution to the heat conductance and $ZT_e$ is the figure of merit of pure electronic origin (obtained assuming $\kappa_{ph}=0$).  Since the high thermoelectric efficiency corresponds to a significant suppression of the electronic thermal conductance near the antibonding state, it is evident that the phonon contribution to the thermal conductance, especially at high temperatures, may play a significant role and can not be neglected. However, it has been shown recently that the phonon thermal conductance may be remarkably reduced in properly prepared silicon nanowire nanojunctions~\cite{hochbaum,boukai,markussenPRL09}, leading to impressive values of $ZT$ ($ZT>1$) at room temperature. The phonon contribution to the heat conductance in quantum dot systems can also be significantly suppressed by appropriate design of the device~\cite{kuoPRB10} (by creating a vacuum layer). Following Ref.~\onlinecite{kuoPRB10}, we have estimated the magnitude of $ZT$ at $k_BT=10\Gamma$ assuming experimentally available values of $\Gamma$,  $\Gamma\approx 2.5$ meV (which is experimentally available in semiconducting quantum dots). Accordingly, for $T\approx 300K$ we find the maximum value of $ZT$ to be of an order of $ZT\approx 18$. To reduce the phonon contribution to heat conductance one can also design the system using phononic crystals as the elements of leads~\cite{clelandPRB01}. Furthermore, one can also minimize the probability of phonon transmission utilizing strong phonon reflection at the interfaces constructed from materials of dissimilar vibrational spectra.~\cite{kimNanoToday} In turn, calculating the phonon contribution to the heat conductance according to Ref.~\onlinecite{Tsaousidou}, one finds $ZT$ up to 8, even in the absence of a vacuum layer.

\begin{acknowledgements}

This work was supported by Polish Ministry of Science and Higher
Education partly as a research project in years 2010 -- 2013 and
partly as a research project in years 2009-2011.

\end{acknowledgements}


\begin{thebibliography}{40}

\bibitem{wied-franz} R. Franz, G. Wiedemann, Ann. Phys. (Berlin)  {\bf 165}, 497 (1853).

\bibitem{mott} M. Cutler and N. F. Mott, Phys. Rev. {\bf 181}, 1336 (1969).

\bibitem{hick} L. D. Hicks and M. S. Dresselhaus, Phys. Rev. B {\bf 47}, 16631 (1993).

\bibitem{beenak} C. W. J. Beenakker and A. A. M. Staring, Phys. Rev. B {\bf 46}, 9667 (1992).

\bibitem{blanter} Y. M. Blanter, C. Bruder, R. Fazio, and H. Schoeller, Phys. Rev. B {\bf 55}, 4069 (1997).

\bibitem{turek} M. Turek and K. A. Matveev, Phys. Rev. B {\bf 65}, 115332 (2002).

\bibitem{koch} J. Koch, F. von Oppen, Y. Oreg, and E. Sela, Phys. Rev. B {\bf 70}, 195107 (2004).

\bibitem{kubala1}B. Kubala and J. K\"onig, Phys. Rev. B {\bf 73}, 195316 (2006).

\bibitem{zianni} X. Zianni, Phys. Rev. B {\bf 75}, 045344 (2007).

\bibitem{zhangxm} X.-M. Zhang, X. Chen, W. Lu, Phys. Lett. A {\bf 372}, 2816 (2008).

\bibitem{kubala2} B. Kubala, J. K\"onig, and J. Pekola, Phys. Rev. Lett. {\bf 100}, 066801 (2008).

\bibitem{murphy} P. Murphy, S. Mukerjee, and J. Moore, Phys. Rev. B {\bf 78}, 161406(R) (2008).

\bibitem{chen} R. Venkatasubramanian, Recent Trends in Thermoelectric Materials Research III, Semiconductors and Semimetals Vol. 71
(Academic Press, New York, 2001), pp. 175–201; G. Chen, Recent Trends in Thermoelectric Materials Research III, Semiconductors
and Semimetals Vol. 71 (Academic, New York, 2001), pp. 203–259.

\bibitem{reddy} P. Reddy, S. Y. Jang, R. A. Segalman, and A. Majumdar, Science {\bf 315}, 1568 (2007).

\bibitem{hochbaum} A. I. Hochbaum, R. Chen, R. D. Delgado, W. Liang, E. C. Garnett,
M. Najarian, A. Majumdar, and P. Yang, Nature (London) {\bf 451}, 163 (2008).

\bibitem{baheti} K. Baheti, J. A. Malen, P. Doak, P. Reddy, S. Y. Jang, T. D.
Tilley, A. Majumdar, and R. A. Segalman, Nano Lett. {\bf 8}, 715 (2008).

\bibitem{boukai} A. I. Boukai, Y. Bunimovich, J. Tahir-Kheli, J. K. Yu, W. A. Goddard III, and J. R. Heath, Nature (London) {\bf 451}, 168 (2008).

\bibitem{schwab} K. Schwab, E. A. Henriksen, J. M. Worlock, and M. L. Roukes, Nature (London) {\bf 404}, 974 (2000).

\bibitem{rego} L. G. C. Rego and G. Kirczenow, Phys. Rev. Lett. {\bf 81}, 232 (1998).

\bibitem{dubi} Y. Dubi and M. Di Ventra, Nano Lett. {\bf 9}, 97 (2009).

\bibitem{swirkowicz} R. \'Swirkowicz, M. Wierzbicki, and J. Barna\'s, Phys. Rev. B {\bf 80}, 195409 (2009).

\bibitem{markussen} T. Markussen, A. P. Jauho, and M. Brandbyge, Phys. Rev. B {\bf 79}, 035415 (2009).

\bibitem{galperin} M. Galperin, A. Nitzan, and M. A. Ratner, Mol. Phys. {\bf 106}, 397 (2008).

\bibitem{lunde} A. M. Lunde, K. Flensberg, and L. I. Glazman, Phys. Rev. Lett. {\bf 97}, 256802 (2006).

\bibitem{segal} D. Segal, Phys. Rev. B {\bf 72}, 165426 (2005).

\bibitem{pauly} F. Pauly, J. K. Viljas, and J. C. Cuevas, Phys. Rev. B {\bf 78}, 035315 (2008).

\bibitem{LiuNano09} Y-S. Liu, Y. -R. Chen, and Y. -C. Chen, ACS Nano {\bf 3}, 3497 (2009).

\bibitem{bergfield} J. P. Bergfield, M. A. Solis, and C. A. Stafford, ACS Nano {\bf 4}, 5314 (2010).

\bibitem{boese} D. Boese and R. Fazio, Europhys. Lett. {\bf 56}, 576 (2001).

\bibitem{dong} B. Dong and X. L. Lei, J. Phys.: Condens. Matter {\bf 14}, 11747 (2002).

\bibitem{krawiec} M. Krawiec and K. I. Wysokinski, Phys. Rev. B {\bf 73}, 075307 (2006).

\bibitem{sakano} R. Sakano, T. Kita, and N. Kawakami, J. Phys. Soc. Jpn. {\bf 76}, 074709 (2007).

\bibitem{kim} T.-S. Kim and S. Hershfield, Phys. Rev. B {\bf 67}, 165313 (2003).

\bibitem{scheibner} R. Scheibner, H. Buhmann, D. Reuter, M. N. Kiselev, and L. W. Molenkamp, Phys. Rev. Lett. {\bf 95}, 176602 (2005).

\bibitem{franco} R. Franco, J. Silva-Valencia, M.S. Figueira, J. Magn. Magn. Mater. {\bf 320}, 242 (2008).

\bibitem{yoshida} M. Yoshida and L. N. Oliveira, Physica B {\bf 404}, 3312 (2009).

\bibitem{costi} T. A. Costi and V. Zlati\'c, Phys. Rev. B {\bf 81}, 235127 (2010).

\bibitem{swirkowiczJP} M. Wierzbicki and R. \'Swirkowicz, J. Phys.: Condens. Matter {\bf 22}, 185302 (2010).

\bibitem{liu} J. Liu, Q.-F Sun, and X. C. Xie, Phys. Rev. B {\bf 81}, 245323 (2010).

\bibitem{kuo} D. M.-T. Kuo, Jpn. J. Appl. Phys. {\bf 48}, 125005 (2009).

\bibitem{hatami} M. Hatami, G. E. W. Bauer, Q. Zhang, and P. J. Kelly, Phys. Rev. B {\bf 79}, 174426 (2009).

\bibitem{Snyder} G. J. Snyder and E. S. Toberer, Nature {\bf 7}, 105 (2008).

\bibitem{wangRev} J.-S. Wang, J. Wang, and J. T. L\"u, Eur. Phys. J. B {\bf 62}, 381 (2008).

\bibitem{dubiRev} Y. Dubi and M. Di Ventra, Rev. Mod. Phys. {\bf 83}, 131 (2011).

\bibitem{zhangdqd} Z.-Y. Zhang, J. Phys.: Condens. Matter {\bf 19}, 086214 (2007).

\bibitem{chiPLA11} F. Chi, J. Zheng, X. -D. Lu, K. -C. Zhang, Phys. Lett. A {\bf 375}, 1352 (2011).

\bibitem{swirkowiczPRB11} M.Wierzbicki and R. \'Swirkowicz, Phys. Rev. B {\bf 84}, 075410 (2011).

\bibitem{orellanaPre11} G. G\'omez-Silva, O. \'Avalos-Ovando, M. L. Ladr\'on de Guevara, and P. A. Orellana, e-print arXiv:1108.4460v2.

\bibitem{trocha07} P. Trocha, J. Barna\'s, Phys. Rev. B {\bf 76}, 165432 (2007).

\bibitem{trochaJNN} P. Trocha and J. Barna\'s, J. Nanosci. Nanotechnol. {\bf 10}, 2489 (2010).

\bibitem{loss} D. Loss and E. V. Sukhorukov, Phys. Rev. Lett. {\bf 84}, 1035 (2000).

\bibitem{guevara} M. L. Ladr\'on de Guevara, F. Claro, P. A. Orellana, Phys. Rev. B {\bf 67}, 195335 (2003).

\bibitem{liuJAP} Y. -S. Liu and X. F. Yang, J. Appl. Phys. {\bf 108}, 023710 (2010).

\bibitem{karlstrom} O. Karlstr\"om, H. Linke, G. Karlstr\"om, A. Wacker, Phys. Rev. B {\bf 84}, 113415 (2011).

\bibitem{uchida} K. Uchida, S. Takahashi, K. Harii, J. Ieda, W. Koshibae,
K. Ando, S. Maekawa, and E. Saitoh, Nature (London) {\bf 455}, 778 (2008);
K. Uchida, T. Ota, K. Harii, S. Takahashi, S. Maekawaa, Y. Fujikawa, E. Saitoh, Solid State Commun.
{\bf 150}, 524 (2010).

\bibitem{johnsonPRB87} M. Johnson, R.H. Silsbee, Phys. Rev. B {\bf 35}, 4959 (1987).

\bibitem{shi} J. Shi, K. Pettit, E. Kita, S.S.P. Parkin, R. Nakatani, M.B. Salamon, Phys. Rev. B {\bf 54}, 15273 (1996).

\bibitem{gravier} L. Gravier, S. Serrano-Guisan, F. Reuse, and J.-P. Ansermet, Phys. Rev. B {\bf 73} 024419 (2006).

\bibitem{gravier2} L. Gravier, S. Serrano-Guisan, F. Reuse, and J.-P. Ansermet, Phys. Rev. B {\bf 73} 052410 (2006).

\bibitem{dubiPRB09} Y. Dubi and M. Di Ventra, Phys. Rev B {\bf 79}, 081302(R) (2009).

\bibitem{LiuJAP11} Y. -S. Liu, F. Chi, X. -F. Yang, and J. -F. Feng, J. Appl. Phys. {\bf 109} 053712 (2011).

\bibitem{rejecPre11} T. Rejec, R. \v{Z}itko, J. Mravlje, and A. Ram\v{s}ak, e-print arXiv:1109.3066v1.

\bibitem{ziman} J. M. Ziman, \emph{Electrons and Phonons. The Theory of Transport Phenomena in Solids} (Oxford University Press, 1960).

\bibitem{wiel} W. G. van der Wiel, S. D. Franceschi, J. M. Elgerman, S. Tarucha, and L. P. Kouwenhoven, Rev. Mod. Phys. {\bf 75}, 1
(2002).

\bibitem{Julliere} M. Julliere, Phys. Lett. A {\bf 54}, 225 (1975).

\bibitem{Stojanovic} N. Stojanovic, D. H. S. Maithripala, J. M. Berg, and M. Holtz, Phys. Rev. B {\bf 82}, 075418 (2010).

\bibitem{markussenPRL09} T. Markussen, A. P. Jauho, and M. Brandbyge, Phys. Rev. Lett. {\bf 103}, 055502 (2009).

\bibitem{kuoPRB10} David M.-T. Kuo and Y.-C. Chang, Phys. Rev. B {\bf 81}, 205321 (2010).

\bibitem{clelandPRB01} A. N. Cleland, D. R. Schmidt, and C. S. Yung, Phys. Rev. B {\bf 64}, 172301 (2001).

\bibitem{kimNanoToday} W. Kim, R. Wang, and A. Majumdar, Nano Today {\bf 2}, 40 (2007).

\bibitem{Tsaousidou} M. Tsaousidou and G. P. Triberis, J. Phys.: Condens. Matter {\bf 22}, 355304 (2010).

\end{thebibliography}
\end{document}